\documentclass[preprint,prd,
superscriptaddress,nofootinbib,showpacs,tightenlines]{revtex4}
\usepackage{amssymb,latexsym}
\usepackage{amsmath,amsbsy}
\usepackage{epsfig,bm}
\usepackage{graphicx}
\usepackage{comment}
\DeclareMathOperator{\tr}{tr}

\DeclareFontFamily{OT1}{pzc}{}
\DeclareFontShape{OT1}{pzc}{m}{it}%
             {<-> s * [1.45] pzcmi7t}{}
\DeclareMathAlphabet{\mathscr}{OT1}{pzc}%
                                 {m}{it}

\begin{document}
\def\a{{\alpha}}
\def\b{{\beta}}
\def\d{{\delta}}
\def\D{{\Delta}}
\def\e{{\varepsilon}}
\def\g{{\gamma}}
\def\G{{\Gamma}}
\def\k{{\kappa}}
\def\l{{\lambda}}
\def\L{{\Lambda}}
\def\m{{\mu}}
\def\n{{\nu}}
\def\o{{\omega}}
\def\O{{\Omega}}
\def\S{{\Sigma}}
\def\s{{\sigma}}
\def\th{{\theta}}

\def\ol#1{{\overline{#1}}}

\def\Dslash{D\hskip-0.65em /}
\def\Dtslash{\tilde{D} \hskip-0.65em /}

\def\CPT{{$\chi$PT}}
\def\QCPT{{Q$\chi$PT}}
\def\PQCPT{{PQ$\chi$PT}}
\def\tr{\text{tr}}
\def\str{\text{str}}
\def\diag{\text{diag}}
\def\order{{\mathcal O}}

\def\meff{{m^2_{\text{eff}}}}

\def\Meff{{M_{\text{eff}}}}
\def\cF{{\mathcal F}}
\def\cS{{\mathcal S}}
\def\cC{{\mathcal C}}
\def\cE{{\mathcal E}}
\def\cB{{\mathcal B}}
\def\cT{{\mathcal T}}
\def\cQ{{\mathcal Q}}
\def\cL{{\mathcal L}}
\def\cO{{\mathcal O}}
\def\cA{{\mathcal A}}
\def\cV{{\mathcal V}}
\def\cR{{\mathcal R}}
\def\cH{{\mathcal H}}
\def\cW{{\mathcal W}}
\def\cM{{\mathcal M}}
\def\cD{{\mathcal D}}
\def\cN{{\mathcal N}}
\def\cP{{\mathcal P}}
\def\cK{{\mathcal K}}
\def\Qt{{\tilde{Q}}}
\def\Dt{{\tilde{D}}}
\def\psit{{\tilde{\psi}}}
\def\St{{\tilde{\Sigma}}}
\def\cBt{{\tilde{\mathcal{B}}}}
\def\cDt{{\tilde{\mathcal{D}}}}
\def\cTt{{\tilde{\mathcal{T}}}}
\def\cMt{{\tilde{\mathcal{M}}}}
\def\At{{\tilde{A}}}
\def\Qt{{\tilde{Q}}}
\def\cNt{{\tilde{\mathcal{N}}}}
\def\cOt{{\tilde{\mathcal{O}}}}
\def\cPt{{\tilde{\mathcal{P}}}}
\def\cI{{\mathcal{I}}}
\def\cJ{{\mathcal{J}}}

\def\eqref#1{{(\ref{#1})}}

\preprint{JLAB-THY-09-965}
\preprint{UMD-40762-446}

\title{Extracting Electric Polarizabilities from Lattice QCD}

\author{W.~Detmold}
\email[]{wdetmold@wm.edu}
\affiliation{%
Department of Physics, 
College of William and Mary, 
Williamsburg, VA 23187-8795,
USA
}
\affiliation{%
Thomas Jefferson National Accelerator Facility, 
Newport News, VA 23606,
USA
}

\author{B.~C.~Tiburzi}
\email[]{bctiburz@umd.edu}
\affiliation{%
Maryland Center for Fundamental Physics, 
Department of Physics, 
University of Maryland, 
College Park,  
MD 20742-4111, 
USA
}

\author{A.~Walker-Loud}
\email[]{walkloud@wm.edu}
\affiliation{%
Department of Physics, 
College of William and Mary, 
Williamsburg, VA 23187-8795,
USA
}

\date{\today}


\begin{abstract}
Charged and neutral, 
pion and kaon electric polarizabilities are 
extracted from lattice QCD using an ensemble of 
anisotropic gauge configurations with dynamical 
clover fermions.
We utilize classical background fields
to access the polarizabilities from two-point correlation functions.
Uniform background fields are achieved by quantizing the electric field strength
with the proper treatment of boundary flux.
These external fields, however, are implemented only in the valence quark sector. 
A novel method to extract charge particle 
polarizabilities is successfully demonstrated for the first time. 
\end{abstract}
\maketitle

%
%

\section{Introduction}

A staple component of electrodynamics courses 
is the electric polarizability.
Neutral materials immersed in electric
fields polarize. 
At the atomic scale, electron clouds distort creating microscopic dipole moments that   
oriente opposite the applied field to minimize the energy.
This simple principle 
accounts for dielectric 
properties of 
materials,
a range of intermolecular forces, 
and 
properties of atoms and nuclei in applied fields.
At the femtoscale, 
hadrons too polarize in applied fields, 
but only 
against the strong chromodynamic interactions
confining their electrically charged quarks into hadrons.

Understanding properties of hadrons 
quantitatively is formidable. 
Quark and gluon interactions
must be treated non-perturbatively for 
which lattice QCD has been developed, 
see~\cite{DeGrand:2006aa} for a review.
Low-energy properties of hadrons, however,  can be described
using an effective theory of QCD, 
based upon treating pseudoscalar mesons
as the Goldstone modes arising from 
spontaneous chiral symmetry breaking.
A picture of hadrons emerges from chiral dynamics: 
that of a hadronic core surrounded by a pseudoscalar meson cloud. 
As some pseudoscalar mesons are charged, 
polarizabilities of hadrons encode
the stiffness of the charged meson cloud (as well as that of the core). 
The form of pseudoscalar meson polarizabilities is consequently
strongly constrained by chiral dynamics%
~\cite{Holstein:1990qy,Burgi:1996qi,Gasser:2006qa}.
Beyond the leading order, 
however, 
results depend on essentially unknown low-energy constants, 
which currently must be estimated in a model-dependent fashion.
For the case of the charged pion, 
confrontation of these results with experiment has proven difficult, 
e.g. from the original measurement%
~\cite{Antipov:1982kz}, 
to the most recent%
~\cite{Ahrens:2004mg},
extracted results disagree with
predictions made using chiral dynamics.
New results with higher statistics and the first kaon results are anticipated from 
COMPASS at CERN~\cite{Abbon:2007pq}.

Lattice gauge theory simulations provide a first principles
approach to determine hadronic polarizabilities from QCD,
crucially test predictions from chiral dynamics,
and
confront experiment.
Indeed the unknown low-energy constants of 
chiral perturbation theory can be determined 
by matching to lattice QCD computations. 
Furthermore, the ability to vary the quark mass
allows one to directly explore the
chiral behavior of observables, 
investigate the convergence properties of the perturbative expansion, 
and thereby test the predictions of the effective theory. 
The highly constrained form for hadronic polarizabilities
within chiral perturbation theory leads to a stringent test 
of low-energy QCD dynamics.

Electric polarizabilities of neutral hadrons 
have been calculated with lattice QCD using the 
quenched approximation at pion masses greater than 
$500 \, \texttt{MeV}$~\cite{Fiebig:1988en,Christensen:2004ca}.
There has also been a fully dynamical calculation of the neutron electric polarizability at a pion mass of 
$760 \, \texttt{MeV}$~\cite{Engelhardt:2007ub}.
These calculations do not employ constant electric fields
but attempt to mitigate effects from field gradients by imposing Dirichlet boundary conditions
on the quark fields in the time and/or space directions.
Such an approach leads to uncertainties that are difficult to quantify.  
In this work, we report on calculations
of pseudoscalar meson polarizabilities using lattice QCD with dynamical configurations.
A salient feature of our computation
is that it utilizes a periodic lattice action with everywhere constant electric fields.
Our calculations of meson polarizabilities 
are the first such to include effects from 
dynamical quarks. At this stage, however, 
we are restricted to electrically neutral sea quarks.
Correcting for this malady would require at least 
an order of magnitude greater computing power.%
\footnote{ 
Pseudoscalar meson polarizabilities first depend on
sea quark charges at next-to-next-to-leading order in the chiral expansion%
~\cite{Hu:2007ts,Tiburzi:2009xx}. 
It is thus possible to extract physical information from simulations
with vanishing sea quark charges by utilizing chiral perturbation theory. 
As the current study is restricted to one volume and one pion mass, 
we leave this investigation to future work.
}
Furthermore, we demonstrate for the first time how to extract 
charge particle polarizabilities from lattice 
two-point correlation functions. 

We begin in Section~\ref{ConstantFields} by describing the implementation
of constant external fields on a lattice.
The Appendix considers the effect of non-uniform fields. 
Next in Section~\ref{Correlators}, we detail how lattice two-point
correlation functions can be utilized
to extract the electric polarizabilities of 
both charged and neutral particles.
Details of our lattice study are then presented in Section~\ref{Results}, 
and summarized in a conclusion, Section~\ref{summy}.

\section{Constant Fields on a Lattice}                                                                    %
\label{ConstantFields}                                                                                              %

To produce a constant electric field, 
$\vec{\cE} = \cE \hat{z}$, 
we use the Euclidean space vector potential,
\begin{equation} \label{eq:A8}
A_\mu(x) = ( 0, 0, - \cE x_4, 0 )
,\end{equation}
where 
$\cE$ 
is a real-valued parameter.
The analytic continuation
$\cE \to -  i \cE_\text{M}$ 
produces a real-valued electric field in Minkowski space. 
Generally this continuation cannot be performed using 
numerical data because of non-perturbative effects,
e.g.~the Schwinger pair-creation mechanism~\cite{Schwinger:1951nm} is absent in Euclidean space.
We are interested, 
however, 
solely in quantities that are perturbative in the external field strength, 
for which  the na\"ive continuation produces the correct Minkowski space physics, 
see~\cite{Tiburzi:2008ma} for explicit details.

To implement the background field on the lattice, 
we modify the 
$SU(3)$ 
color gauge links,
$U_\mu(x)$, 
for each quark flavor by multiplying by the color-singlet Abelian links,
$U_\mu^{(\cE)} (x)$,
for the external field, 
namely
\begin{equation}  \label{eq:AbelianLink}
U_\mu (x) 
\longrightarrow 
U_\mu (x) U_\mu^{(\cE)} (x)
,\end{equation}
where
$U_\mu^{(\cE)} (x) = \exp [ i Q A_\mu(x) ]$,
where 
$Q$
is the quark electric charge, 
and 
$A_\mu(x)$
is given in Eq.~\eqref{eq:A8}.
As this multiplication is carried out on pre-existing gauge configurations, 
the sea quarks remain electrically neutral. 
This approximation is imposed because of computational restrictions which will not 
be rectified in the near future without a significant increase in resources.

The inclusion of the field via Eq.~\eqref{eq:AbelianLink}
does not lead to a constant electric field.
On a torus, 
constant gauge fields require quantization~\cite{'tHooft:1979uj,'tHooft:1981sz,vanBaal:1982ag}. 
The basic argument is as follows. 
With periodic boundary conditions,%
\footnote{%
The argument applies equally well to the case
of twisted boundary conditions on the matter fields $\psi(x)$ of the form: 
$\psi ( x + L \hat{x}_j ) = e^{i \theta_j} \psi(x)$,
and analogously for the time direction.
Dirichlet boundary conditions, on the other hand, inevitably lead to problems.
}
the action is defined on a  torus,
which is a closed surface. 
For the field we wish to implement, 
the only plane with non-vanishing flux 
is the $x_3$-$x_4$ plane. 
The total area of the 
$x_3$-$x_4$ plane is
$\beta L$,
where $L$ is the length of the $x_3$-direction, 
and $\beta$ is the length of the $x_4$-direction.  
Because the torus is a closed surface, however, 
there can be no net flux through the 
$x_3$-$x_4$ plane (modulo $2 \pi$),
i.e.~%
$\Phi =  Q  \cE \beta L \equiv 2 \pi n $, 
with $n$ as an integer. 
This leads to the 't Hooft quantization condition
\begin{equation} \label{eq:Quant}
\cE 
= 
\frac{ 2 \pi n} { q_d \, \beta L}
.\end{equation}
Here we have used the down quark electric charge, 
$q_d = - 1/3 \, e$,  
and note that the up quark will necessarily encounter properly quantized
fields when Eq.~\eqref{eq:Quant} is met because 
$q_u = - 2 q_d$.

The argument presented for constant gauge fields applies to a continuous torus, 
and must be modified for a discrete torus, 
see e.g.~\cite{Smit:1986fn,Rubinstein:1995hc,AlHashimi:2008hr}. 
On a discrete torus, 
each of the elementary plaquettes must be identical with value:
$\exp ( i Q \cE )$ 
to arrive at the constant electric field 
$\cE$.
With Eq.~\eqref{eq:AbelianLink}, 
the plaquettes are identical in the bulk of the lattice but not at the boundary, 
where there are $L$ plaquettes with differing flux.
Each of these plaquettes wraps around from $x_4 = \beta - 1$ to $x_4 = 0$, 
with the common value:
$\exp [ i Q \cE ( 1 - \beta)]$.
This unwanted flux can be eliminated on $L-1$ of the plaquettes
by including additional transverse links, 
$U_\mu^{(\cE)_\perp}(x)$, 
at the boundary, 
\begin{equation} \label{eq:ModifiedAbelianLink}
U_\mu (x) 
\longrightarrow
U_\mu (x) U_\mu^{(\cE)} (x)
U_\mu^{(\cE)_\perp}(x)
,\end{equation} 
with 
$U_\mu^{(\cE)_\perp}(x)
= 
\exp [ i Q \cE \beta x_3 \, \delta_{\mu 4} \, \delta_{x_4, \beta - 1} ]$.
Now the field through every plaquette is $\cE$,
with only one exception: 
the plaquette at the far corner of the lattice,
$( x_3, x_4 ) = ( L-1, \b - 1)$, 
that wraps around to $x_3 = 0$, and $x_4 = 0$. 
The value of this plaquette is: 
$\exp [ i Q \cE ( 1 - \beta L)]$, 
which is identical to the plaquette in the bulk of the lattice  
provided 't Hooft's quantization condition, Eq.~\eqref{eq:Quant}, is met.
We have previously demonstrated the effects of using non-quantized field values, 
finding non-negligible shifts in particle spectra~\cite{Detmold:2008xk}. 
We summarize our findings in the Appendix. 
In this work, we implement the external field using Eq.~\eqref{eq:ModifiedAbelianLink},
and quantized values for the field strength $\cE$ in Eq.~\eqref{eq:Quant}. 
This choice corresponds to a completely periodic lattice gauge action,
and thus corresponds to a field theory at  a finite (but low) temperature
in the continuum limit.

\section{Correlation Functions}                                                                              %
\label{Correlators}                                                                                                     %

For a neutral particle, it is straightforward to 
calculate the electric polarizability using 
standard lattice spectroscopy~\cite{Fiebig:1988en}.
\footnote{%
Strictly speaking this is only true at infinite volume. 
At finite volume, there are additional effects stemming 
from boundary conditions and the compact nature of the 
external gauge field~\cite{Hu:2007eb,Tiburzi:2008pa}. 
As we employ only one lattice volume to demonstrate our methods, 
we neglect these additional corrections at this stage.
Further analysis with multiple volumes and multiple pion masses 
is needed to control these systematics. 
}
One merely matches the long-time behavior of Euclidean two-point functions, 
$g(x_4, \cE)$,
computed in QCD to the expectations of the effective hadronic theory 
to deduce the particle's energy.
The lattice two-point function has the form
\begin{equation}
g(x_4, \cE) 
= 
 \sum_{\bm{x}} \, \langle  0 | \phi(x) \phi^\dagger(0) | 0 \rangle_{\cE}
,\end{equation}
where 
$\phi$
is an interpolating field for the particle of interest
(e.g. 
$\phi = \ol d \gamma_5 s$
for the 
$K^0$),
and the subscript 
$\cE$ 
denotes that the correlation function is determined in the background electric field. 
This correlation function is matched onto the correlator 
$G(x_4, \cE)$, 
in the hadronic theory, 
\begin{eqnarray} 
G(x_4, \cE) &=& 
Z(\cE) e^{- E(\cE) x_4} +
Z'(\cE) e^{- E'(\cE) x_4} 
+ \ldots
\label{eq:falloff}
,\end{eqnarray}
where the ellipsis represents exponentially suppressed contributions
beyond the first excited state. 
The ground-state particle's energy, 
$E(\cE)$, 
has a series expansion in the 
external field strength
\begin{equation} \label{eq:energy}
E(\cE) = 
M 
+ 
\frac{1}{2} 4 \pi \alpha_E \cE^2 
- 
\frac{1}{4!} (4 \pi)^2 \ol \alpha_{EEE} \cE^4
+ \ldots
,\end{equation}
where 
$M$ 
is the particle's mass,
$\alpha_E$ 
its electric polarizability, 
and
$\ol \alpha_{EEE}$
is a multiple electric dipole
interaction strength. 
Here the ellipsis represents terms at higher order 
in the strength of the field. 
The sign of the polarizability term (quadratic Stark shift) 
is positive due to our treatment in Euclidean space.
The amplitudes, 
$Z(\cE)$ 
and 
$Z'(\cE)$, 
also have expansions in even powers of 
$\cE$. 
As explained below, 
we are forced in our particular computations to consider
contributions from excited states, shown in Eq.~\eqref{eq:falloff}.
The energy, $E'$, of the first excited state has an analogous weak field expansion
in terms of the mass $M'$, polarizability $\alpha'_E$, \emph{etc}.

When charged particles are subjected 
to constant electric fields, 
we again match the lattice 
correlation function, 
$g(x_4, \cE)$,
to the correlator calculated in the hadronic theory.
With sufficiently weak fields, quarks and gluons will still hadronize into a tower of states
of the same quantum numbers, specifically of the same charge. 
For times, 
$x_4$, 
long compared to that set by the 
excited state mass,
$x_4 \gg \tau' \sim  1 / M'$,
the excited state contributions to the 
two-point function will still be exponentially suppressed
(albeit not a simple exponential).
For times beyond $\tau'$, 
we can assume the two-point correlation function 
will be dominated by the ground state. 
As this state is charged, the behavior of the correlation 
function will have a more complicated form than 
a simple exponential falloff with time.

For a relativistic scalar particle of charge $Q$, 
consider the single-particle effective action in the hadronic theory.
As the particle is composite, there are both Born
and non-Born terms in the action. The non-Born 
terms account for non-minimal couplings
of the field to the particle, 
such as polarizabilities. 
These couplings can be summed, 
as in the case of a neutral particle, 
into the energy, 
$E(\cE)$, 
defined above. 
The Born couplings additionally must be summed to 
arrive at the charged particle two-point function. 
For the field specified by Eq.~\eqref{eq:A8}, 
the equations of motion for a scalar particle are exactly solvable, 
and lead to the two-point function in the hadronic theory
\begin{eqnarray} \label{eq:GCharged}
G(x_4, \cE)
&=&
Z(\cE) \, D\big(x_4, E(\cE), \cE \big)
+
Z'(\cE) \, D\big(x_4, E'(\cE), \cE \big)
+ 
\ldots
,\end{eqnarray} 
with the ellipsis representing contributions beyond the first excited state, 
and the relativistic propagator of a charged scalar, 
$D\big(x_4, E(\cE), \cE\big)$, 
given by~\cite{Tiburzi:2008ma}
\begin{equation} \label{eq:QProp}
D\big(x_4, E(\cE), \cE\big)
=
\int_0^\infty ds
\sqrt{ \frac{ Q \cE }{2 \pi \sinh ( Q \cE  s ) } }
\exp
\left[
- \frac{ Q \cE  x_4^2}{2} \coth (  Q \cE   s )
- \frac{E(\cE)^2 s}{2}
\right]
,\end{equation}
where 
$E(\cE)$ 
can no longer be interpreted as the energy
but remains given by Eq.~\eqref{eq:energy}.
Classically 
$E(\cE)$ 
is the rest energy of the charged particle. 
The analogy to a classically accelerating particle
occurs at the level of the particle's action.
Interpreting the Euclidean time behavior 
of the particle's motion on a compact space 
in terms of acceleration proves difficult. 
For 
$Q=0$,
this propagator properly reduces to
Eq.~\eqref{eq:falloff}. 
For sufficiently weak fields, or equivalently
short times, the $\cO(\cE^2)$ term in the
series expansion of the correlator
reproduces the non-relativistic result
derived in~\cite{Detmold:2006vu}.
Due to our particular anisotropic lattices, 
we include contributions to the two-point function 
from the first excited state thereby stabilizing the extraction
of ground state parameters. 
The quantum numbers of the excited state are identical to the 
ground state, 
i.e.~$Q' = Q$.

\section{Lattice Results}                                                                                              %
\label{Results}                                                                                                               %

To demonstrate our method for extracting meson 
polarizabilities from lattice two-point functions, 
we have employed an ensemble of anisotropic gauge
configurations with $2+1$-flavors of dynamical clover fermions~\cite{Edwards:2008ja,Lin:2008pr}.
The ensemble we use consists of 
$200$ 
lattices of size 
$20^3 \times 128$. 
After an initial 
$1000$ 
thermalization trajectories, 
the lattices were chosen 
from an ensemble of 
$7000$
spaced either by 
$20$ 
or
$40$
to minimize autocorrelations. 
The spatial lattice spacing of these configurations is
$a_s = 0.123 \, \texttt{fm}$~\cite{Edwards:2008ja,Lin:2008pr},
with a non-perturbatively tuned 
anisotropy parameter of 
$\xi \equiv a_s / a_t = 3.5$,
where 
$a_t$ 
is the temporal lattice spacing. 
The finer temporal spacing affords us 
the ability to better fit non-standard behavior 
for two-point correlation functions, and is critical 
for this analysis. 
On the ensemble, 
the renormalized strange mass 
is near the physical value, 
while the renormalized light quark mass 
leads to a pion mass of
$m_\pi \approx 390\, \texttt{MeV}$.

On each configuration, 
we compute at least 
$10$ 
up quark propagators, 
$10$ 
down quark propagators, 
and 
$10$ 
strange quark propagators with random spatial source locations.
Multiple inversions were made efficient using the EigCG inverter%
~\cite{Stathopoulos:2007zi}. 
Interpolating fields at the source are generated from gauge-covariantly
Gaussian-smeared quark fields~\cite{Teper:1987wt,Albanese:1987ds} 
on a stout-smeared~\cite{Morningstar:2003gk}
gauge field in order to optimize the overlap onto the ground state. 
Interpolating fields at the sink are constructed from local quark fields.
Each propagator is located with source time at 
$\tau_{\text{src}} = 0$.  
Randomization of the source time location, 
while improving the statistical sampling, 
would complicate the extraction of both charged and neutral meson correlation functions, 
as two-point functions are no longer time-translationally invariant. 
For charged particles, the correlator in Eq.~\eqref{eq:GCharged} 
is explicitly a function of the sink time-slice and not simply 
a function of the source-sink separation
(the full dependence on source time is given in~\cite{Tiburzi:2008ma}), 
while for neutral particles the violation of time-translation invariance arises from volume effects.

The external field was implemented using Eq.~\eqref{eq:ModifiedAbelianLink}, 
and propagators were computed for nine values of the field strength, $n$,
corresponding to the integer appearing in the quantization condition, Eq.~\eqref{eq:Quant}. 
We use 
$n=0$, 
which corresponds to a vanishing external field, 
as well as 
$n = \pm 1$, $\ldots$, $ \pm 4$. 
On our lattices, 
the expansion parameter governing the deformation of a hadron's pion cloud is given by~\cite{Tiburzi:2008ma}
\begin{equation}
\left( \frac{e \, \cE}{m_\pi^2} \right)^2
= 
0.18 \, n^2
.\end{equation} 
From the size of this parameter, 
we anticipate the need to include terms beyond quadratic order in the electric field expansion of hadron energies. 
In our analysis, 
we include terms up to quartic order. 
Larger lattices will be required for better control over systematics relating to the electric field expansion of observables.

Meson two-point functions were obtained 
for each source location on a given configuration. 
Individual results for the multiple source locations on each configuration
were then source averaged. 
This procedure was carried out for each value of
the external field. 
To satisfy invariance under parity transformations, 
under which $\cE \to - \cE$, 
we took the geometric mean of correlators calculated 
at 
$n$, 
and 
$-n$
on each configuration. 
This reduced the set of fields to five magnitudes corresponding to the integers, 
$n=0, \ldots, 4$. 
The ensemble of correlation functions was then used to generate
$200$ 
bootstrap ensembles. 
Fits to the bootstrapped ensemble were performed as described below.%
\footnote{
We have performed multiple differing procedures to analyze the data, 
of which only one is described in detail in the text. Throughout we will 
comment on the alternate procedures. 
Most notably, fits have been performed using a jackknife procedure
to determine uncertainties, and with a separate analysis of the positive 
and negative values of the field strength. Effects of correlations in the 
data have been investigated by blocking neighboring configurations, 
and consistent results have been obtained.
}

\subsection{Neutral pion and kaon}

\begin{figure}
\epsfig{file=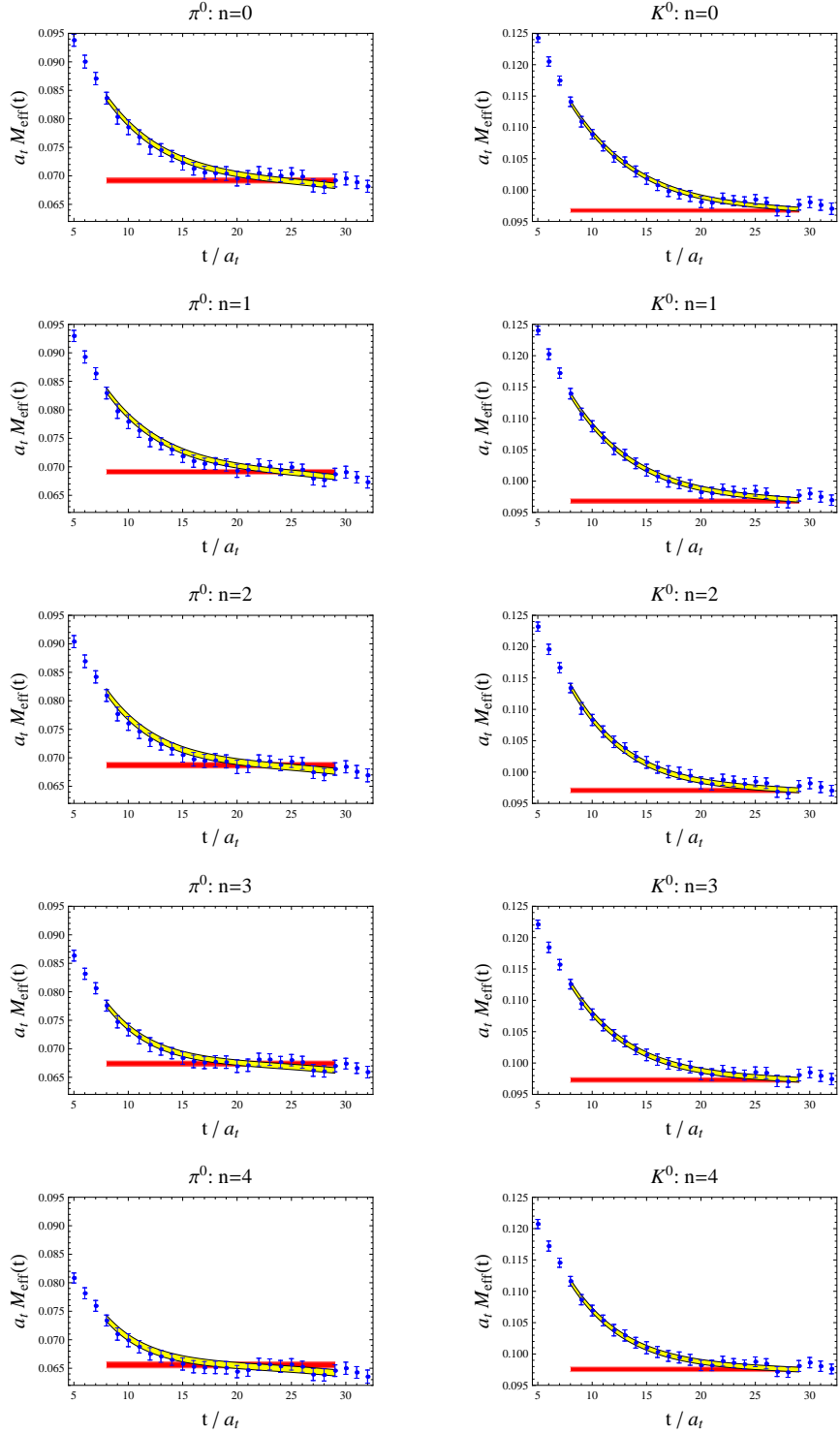,width=0.75\textwidth}
\caption{
Effective mass plots for the (connected) neutral pion and neutral kaon. 
Values for $n$ correspond to the magnitude of the quantized electric field $\cE$, 
Eq.~\eqref{eq:Quant}. 
Curved bands show fits to the correlation functions. 
For each plot, the band spans the fit window, 
and the width is set by the uncertainty in the ground state energy. 
The flat bands shown are ground state energies with uncertainty for each field value.
On a lattice of infinite time extent,  
the effective masses should asymptote to these bands over long times. 
}
\label{f:PiZeroMeff}
\end{figure}

Bootstrapped correlators for the neutral pion and kaon were obtained using the procedure described above. 
As we use standard spectroscopy to determine the polarizabilities for neutral particles, 
we handle these mesons first. 
To facilitate the discussion, 
we consider the standard effective mass, given by
\begin{equation} \label{eq:Meff}
M_{\text{eff}}(t) =  - \log \frac{g(t + 1,\cE)}{g(t,\cE)}
,\end{equation}
where $g(t,\cE)$ is the bootstrap ensemble-averaged correlator.
Error bars on the effective mass are determined
using the bootstrap ensemble. 
Effective mass plots for the neutral pion and neutral kaon
are shown in Fig.~\ref{f:PiZeroMeff}. 
For the neutral pion, so far we have  only calculated the connected part of the correlation function.
The effective mass plot should exhibit a plateau over a range of time when the 
ground state saturates the correlation function. 
The temporal extent of our anisotropic lattices is 
$\beta = 128 \, a_t =  4.5 \, \texttt{fm}$, 
which is considerably smaller than typical isotropic lattices, 
where 
$\beta > 7.5 \, \texttt{fm}$. 
As one must wait long enough for the excited states to drop out, 
the pion and kaon effective masses never plateau (or barely exhibit a plateau) 
because of the backward propagating image from the time boundary.

To extract the ground state properties, 
we fit the correlation function 
$g(t,\cE)$ 
using the two-state form of 
$G(t,\cE)$ 
in Eq.~\eqref{eq:falloff}
augmented to include a backwards propagating ground state,
and backwards propagating excited state
(i.e.~a  sum of two hyperbolic cosines).
We use a correlated chi-squared analysis
to fit the time-dependence of the bootstrap ensemble of correlators. 
To determine the fit window, 
we use black box methods comparing single and double effective masses, 
see~\cite{Fleming:2004hs,Fleming:2009wb,Beane:2009ky} for details on the latter.
We found the same fit window, 
$8 \leq t \leq  29$,  
could be used for a given particle for every value of the field strength. 
Alternate fits on the same window without backwards propagating states
result in a $1.5\%$ shift of the neutral pion energies, 
and a negligible shift of the kaon energies.

As the parameters 
$Z(\cE)$ 
and 
$Z'(\cE)$ 
enter the fit function 
$G(t,\cE)$ 
linearly, 
we utilize variable projection (see~\cite{Fleming:2004hs} for references)
to reduce the number of fit parameters from four down to two, 
namely just the energies
$E(\cE)$ 
and 
$E'(\cE)$.%
\footnote{
We also analyze correlation functions by fitting the effective masses 
with two states. These three parameter fits give consistent results. 
} 
We perform these two-state fits on the entire bootstrap ensemble arriving
at an ensemble of energies for each magnitude of the electric field $\cE$, 
in particular
$\{ \mathscr{E}_i(\cE) \}$ 
for the ground state, 
where $i$ indexes the bootstrap sample, $i = 1, \ldots, N$. 
As the ensembles of configurations for different field strengths are generated
from the same underlying lattice configurations, 
correlations between the energies for different field strengths
will be significant and it is important to account for these.
On the bootstrap ensemble of energies, 
we perform electric-field correlated fits  
to the energy function 
$E(\cE)$
given in Eq.~\eqref{eq:energy}. 
With the ensemble average energies denoted by 
$\mathscr{E}(\cE) = \frac{1}{N} \sum_i \mathscr{E}_i(\cE)$, 
we minimize the correlated chi-squared, namely 
\begin{equation}
\chi^2  \label{eq:Ecorr}
= 
\sum_{\cE, \cE'}
\Big[
\mathscr{E}(\cE) - E( \cE)
\Big]
C^{-1}_{\cE, \cE'}
\Big[
\mathscr{E}(\cE') - E( \cE')
\Big]
,\end{equation}
with the field-strength correlation matrix, 
$C_{\cE,\cE'}$, 
given by
\begin{equation}
C_{\cE, \cE'} 
= 
\frac{1}{N -1}
\sum_{i = 1}^{N}
\, \Big[\mathscr{E}(\cE) -  \mathscr{E}_i (\cE) \Big] \,\Big[  \mathscr{E}(\cE') -  \mathscr{E}_i(\cE') \Big]
.\end{equation}
Because all three fit parameters, 
$M$, $\alpha_E$, and $\ol \alpha_{EEE}$,
enter the fit function 
$E(\cE)$ 
linearly, 
the chi-squared minimization can be done analytically.  
Fits to the energy function are carried out on the bootstrap ensemble, 
resulting fit parameters are averaged, and the uncertainties from 
fitting and bootstrapping are added in quadrature. 
We perform two different field-correlated fits as follows:
(I) a fit to all five field strengths using Eq.~\eqref{eq:energy}, 
(II) the same fit function but with the largest field strength excluded.
Finally, to estimate the systematics due to the choice of fit window, 
we performed uncorrelated fits to the electric field dependence 
of meson energies determined on adjacent fit windows. 
We chose the nine fit windows obtained by varying the start and end times
by one unit in either direction. 
On each time window, we determined the electric polarizability.
The systematic uncertainty on 
$\alpha_E$ 
due to the fit window is estimated as the standard deviation of the extracted 
$\alpha_E$ 
over the various adjacent windows.
Fit details and extracted parameters are tabulated in Table~\ref{t:NeutralFit}.

%
\begin{table}[t]
\begin{center}
\begin{tabular}{cccc||cccc}
$\pi^0$ & $\quad n \quad $ & $a_t E(\cE)$ & $\quad 1-P \quad $ & $K^0$ & $\quad n \quad$ & $a_t E(\cE)$ & $\quad 1-P \quad$ 
\tabularnewline
\hline
\hline
$$ & $0$ & $0.0692(5)$ & $0.90$ &
$$  & $0$ & $0.0967(3) $ & $0.89$ 
\tabularnewline
& $1$ & $0.0691(5) $ &$0.70$ &
& $1$ & $0.0968(4)$ & $0.87$
\tabularnewline
& $2$ & $0.0687(5) $ & $0.76$ &
& $2$ & $0.0971(4)$ & $0.84$
\tabularnewline
& $3$ & $0.0674(5) $ & $0.74$ &
& $3$ & $0.0973(4) $  & $0.92$
\tabularnewline
& $4$ & $0.0656(5) $ & $0.83$ &
& $4$ & $0.0976(4) $ & $0.94$ 
\tabularnewline
\hline
\hline
\tabularnewline
\end{tabular}
\begin{tabular}{ccccc||ccccc}
$\pi^0 $ & $\quad  a_t M \quad $ &  $\quad \alpha_E^{\text{latt}} \quad  $  & $\quad \ol \a^{\text{latt}}_{EEE} \quad $ & $1-P $ &
$K^0 $ & $\quad a_t M \quad $ & $\quad \alpha_E^{\text{latt}} \quad  $   & $\quad \ol \a^{\text{latt}}_{EEE} \quad $ & $ 1-P $
\tabularnewline
\hline
I & $0.0692(1)$ & $-2.6(5)(9)$ & $1.8(5)$ & $0.69$ &
I & $0.0968(1)$ & $\phantom{-}1.5(4)(7)$ & $0.6(5)$ &  $0.97$
\tabularnewline
II & $0.0692(1)$ & $-1.0(1.5)(1.4)$ & $5.3(3.2)$ & $0.92$ &
II & $0.0967(1)$ & $\phantom{-}1.8(1.0)(1.9)$ & $1.3(1.9)$ &  $0.95$
\tabularnewline
\hline
\hline
\end{tabular}
\end{center}
\caption{%
Summary of fit results for neutral meson two-point functions for 
$8 \leq t \leq  29$.
Here $\pi^0$ refers to the connected part of the correlation function, 
the fine-structure constant is
$\a_{f.s.} = e^2 / 4 \pi$,
and 
$1-P$ is the integrated chi-squared.
All quoted values are averages over the bootstrap ensemble, and are given in dimensionless lattice units.
For the electric polarizability, $\alpha_E^{\text{latt}} = \alpha_E (2 \a_{f.s.} a_t a_s^2)^{-1}$
and the higher-order coupling, $\ol \a^{\text{latt}}_{EEE} = \ol \alpha_{EEE} ( 4! \, \a_{f.s.}^2 a_t^3 a_s^4)^{-1} 10^{-3} $.
The first half of the table summarizes the time-correlated fits to the energies in each field, 
while the second half summarizes the field-correlated fits. 
The fits I and II are described in the text. 
The second uncertainty on the polarizabilities is an estimate
of the systematic due to the choice of fit window as explained in the text. 
}
\label{t:NeutralFit}
\end{table}

%
%
%
%
%
%
\begin{figure}[!t]
\epsfig{file=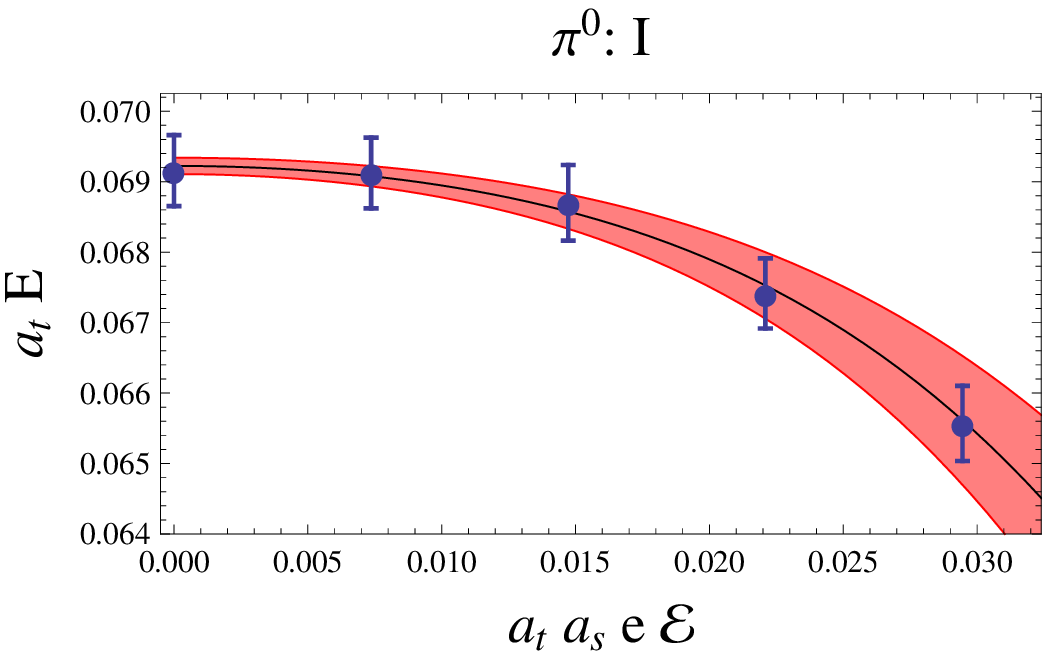,width=7.5cm}
\epsfig{file=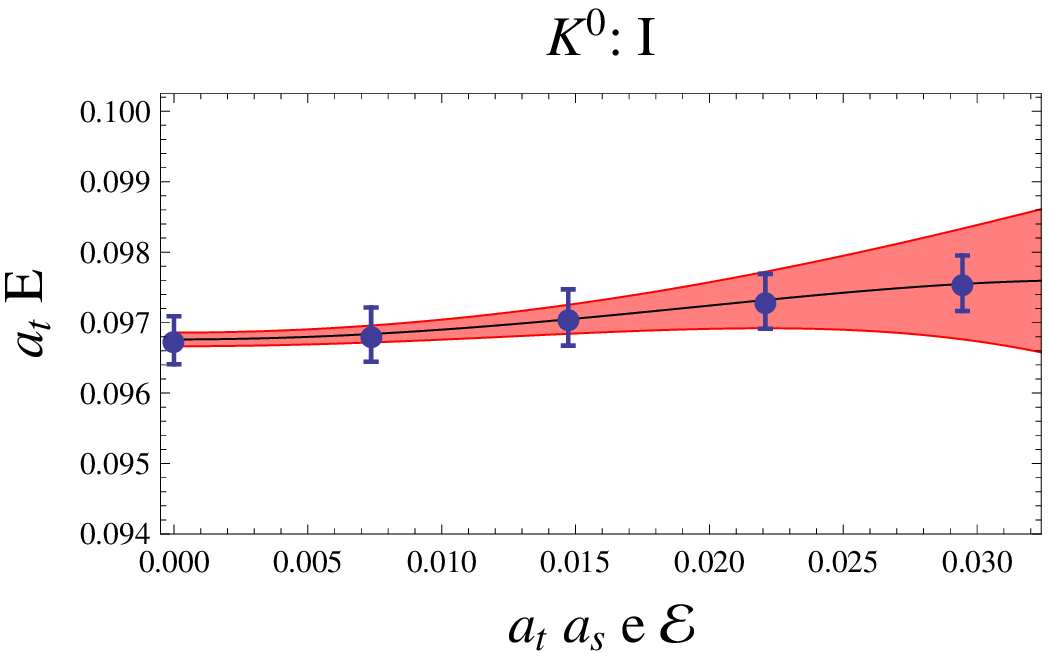,width=7.5cm}
\\
\epsfig{file=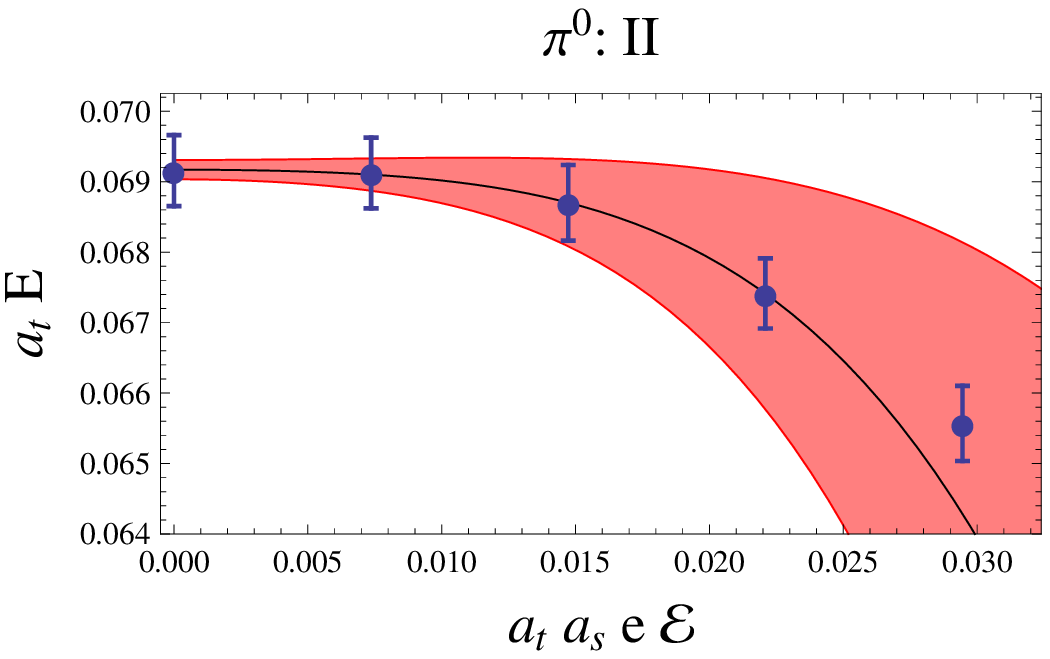,width=7.5cm}
\epsfig{file=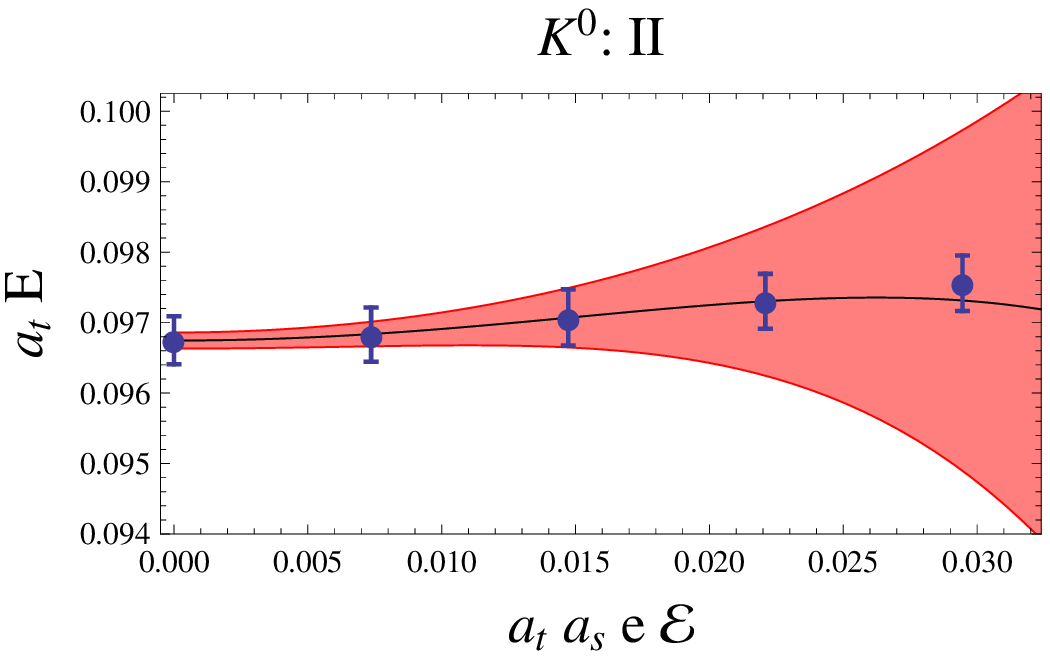,width=7.5cm}
\caption{
Plots of correlated fits to the electric field dependence of neutral meson energies.
For each field strength, 
the bootstrap averaged energies are plotted with
error bars reflecting the uncertainty from statistics and fitting. 
Fits I and II to the $\cE$-dependence are also shown with the plotted bands reflecting the uncertainty
in the parameters appearing in Eq.~\eqref{eq:energy}. 
}
\label{f:NeutralEvsE}
\end{figure}

From the extracted polarizabilities,  
we can investigate the electric field dependence of meson energies. 
This is done in Fig.~\ref{f:NeutralEvsE} for the neutral pion and neutral kaon. 
For the connected part of the neutral pion, we see downward curvature 
of the energy with respect to increasing 
$\cE$, 
while for the neutral kaon the energy is comparatively quite flat. 
In physical units, the polarizabilities 
$\a_E^{\pi^0}$, 
and
$\a_E^{K^0}$, 
are not consistent with na\"ive expectations. 
To attempt a qualitative explanation for the size of the ground state polarizabilities, 
we compare our results with predictions from chiral perturbation theory. 
The neutral pion electric polarizability at one-loop is negative~\cite{Bijnens:1987dc,Donoghue:1988ee}. 
While this is surprising, the one-loop polarizability arises solely 
from the disconnected contraction between quark basis 
$\eta_u$ and $\eta_d$ mesons~\cite{Hu:2007ts}.
Hence the negative sign owes to group theory weight of 
$\eta_u$ 
versus 
$\eta_d$ 
in the pion interpolating field,
$\pi^0 \sim \frac{1}{\sqrt{2}} ( \eta_u - \eta_d)$. 
As we have only calculated the connected part of the correlator,
chiral perturbation theory suggests that
$\a_E^{\pi^0}$ 
is an order of magnitude
smaller than the na\"ive expectation. 
While our result is of this magnitude, it is of the wrong sign
(the average of $\eta_u$ and $\eta_d$ polarizabilities should be positive). 
This negative value could arise from volume effects, 
which are known to be non-vanishing at next-to-leading order
in chiral perturbation theory~\cite{Tiburzi:2008pa}. 
For the neutral kaon polarizability, 
the one-loop chiral computation vanishes, 
even with electrically neutral sea quarks~\cite{Tiburzi:2009xx}. 
Our extracted neutral kaon polarizability, however, is smaller 
than typical two-loop contributions. 
Because the dominant volume corrections arise from pion loops, 
we expect the neutral pion and kaon volume effects to be of the same size. 
If the negative result for the connected $\pi^0$ is due to volume corrections, 
then the near vanishing result for the $K^0$ could be due to a near cancelation
between the polarizability and the volume effect. 
Further study at multiple volumes and pion masses is 
necessary to disentangle the chiral and volume corrections.

\subsection{Charged pion and kaon}

\begin{figure}
\epsfig{file=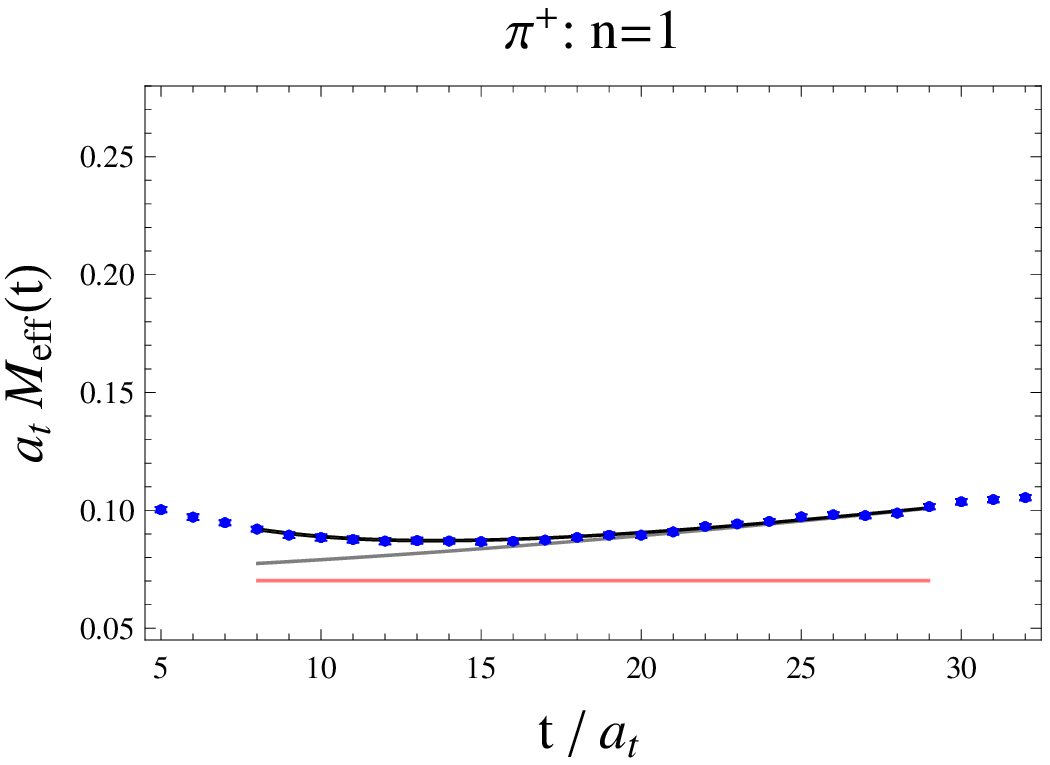,width=6.65cm}
$\phantom{sp}$
\epsfig{file=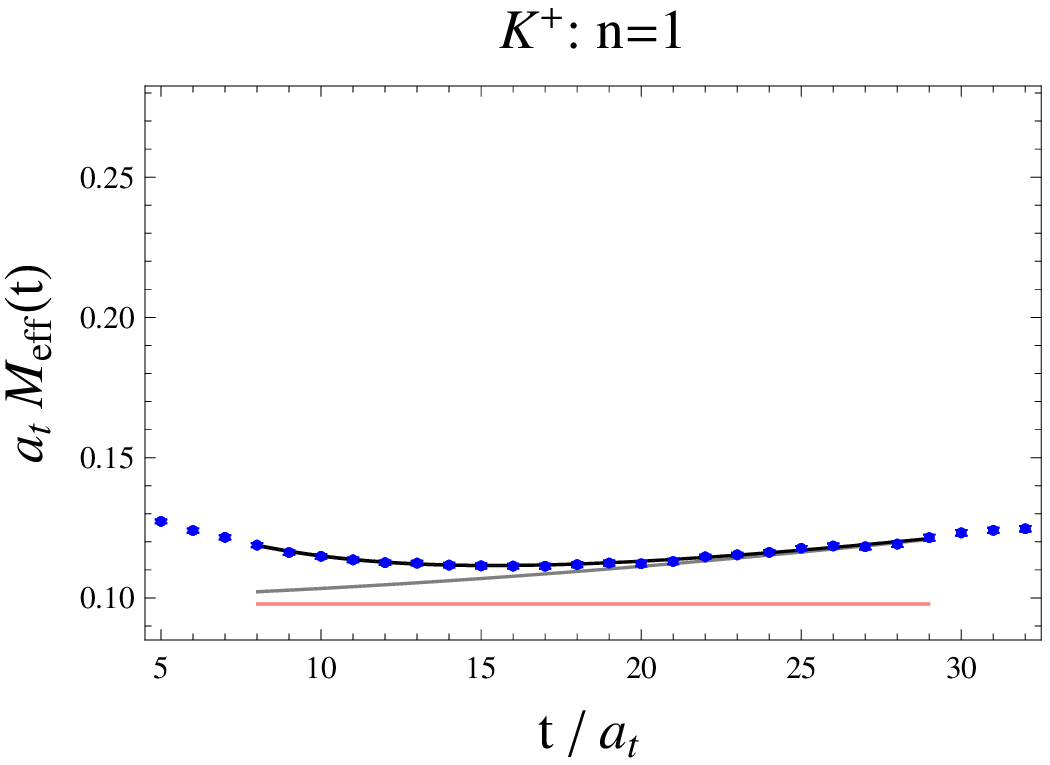,width=6.65cm}
\\
\epsfig{file=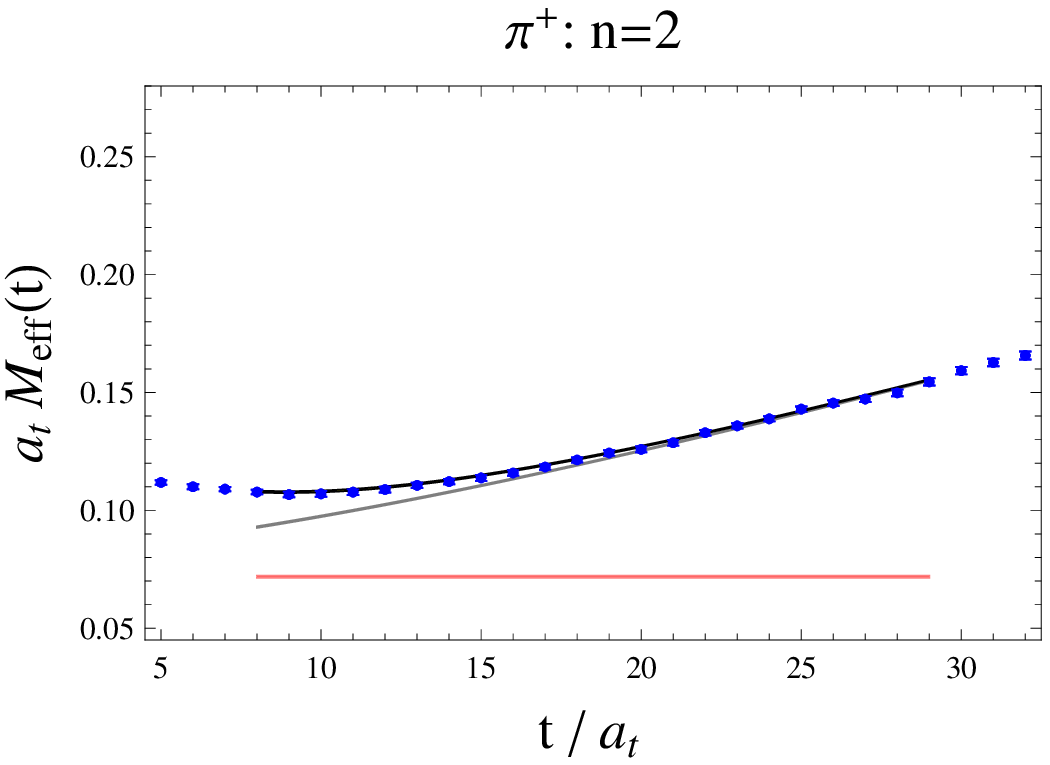,width=6.65cm}
$\phantom{sp}$
\epsfig{file=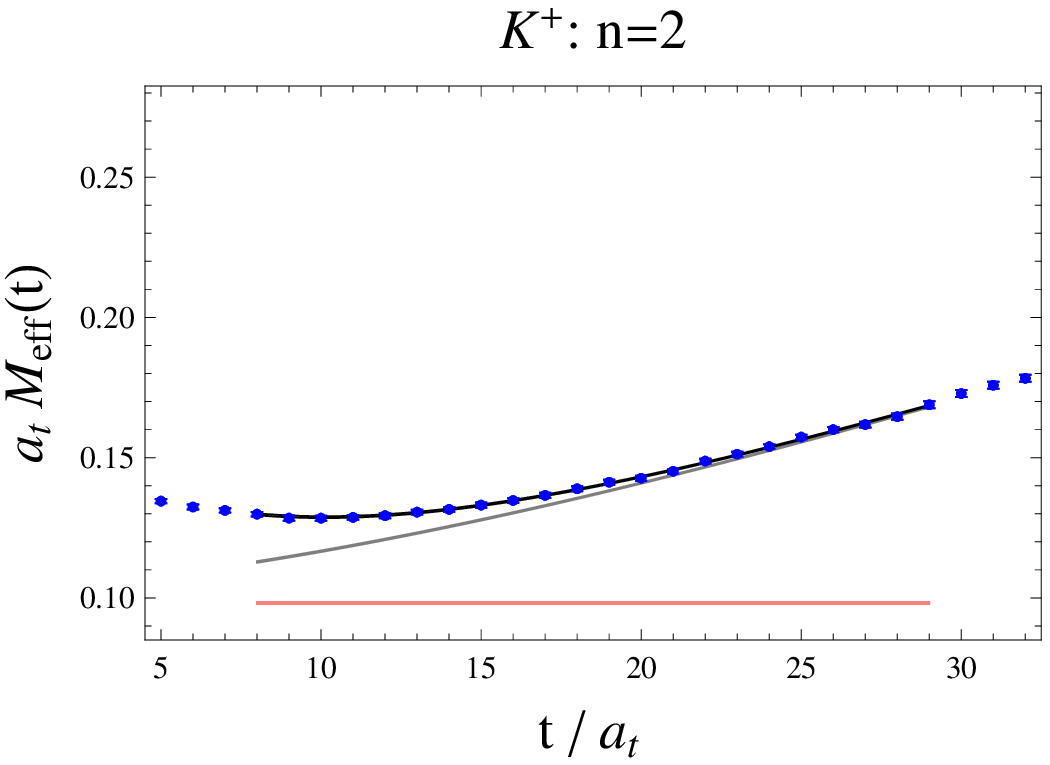,width=6.65cm}
\\
\epsfig{file=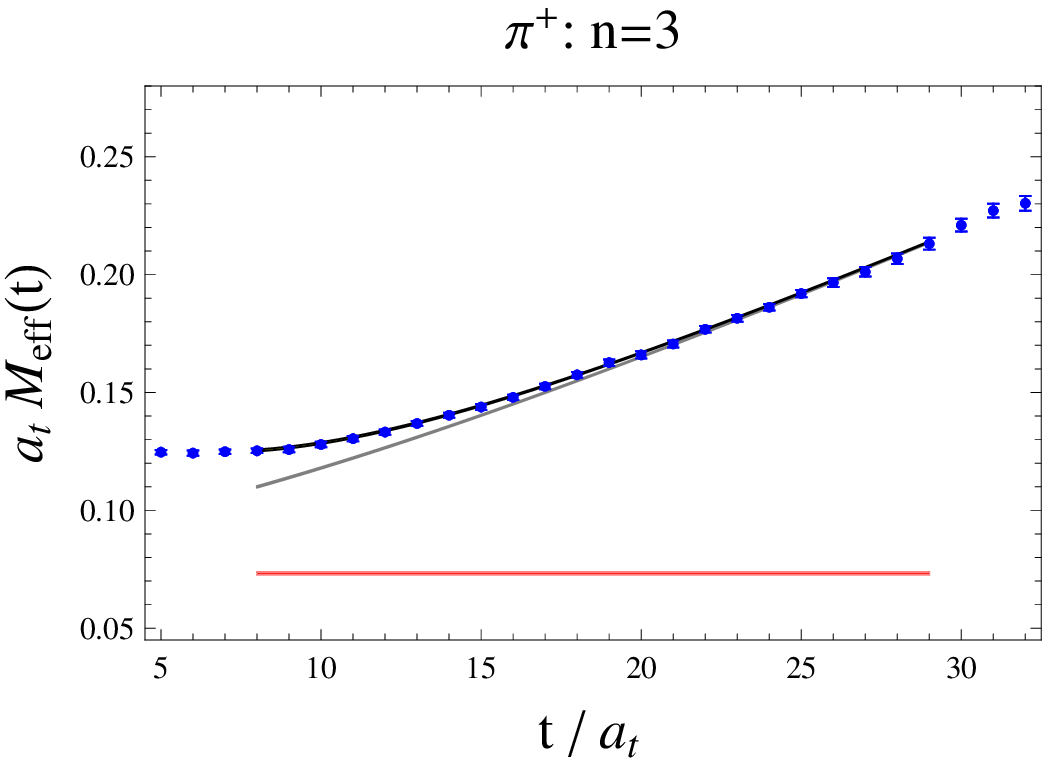,width=6.65cm}
$\phantom{sp}$
\epsfig{file=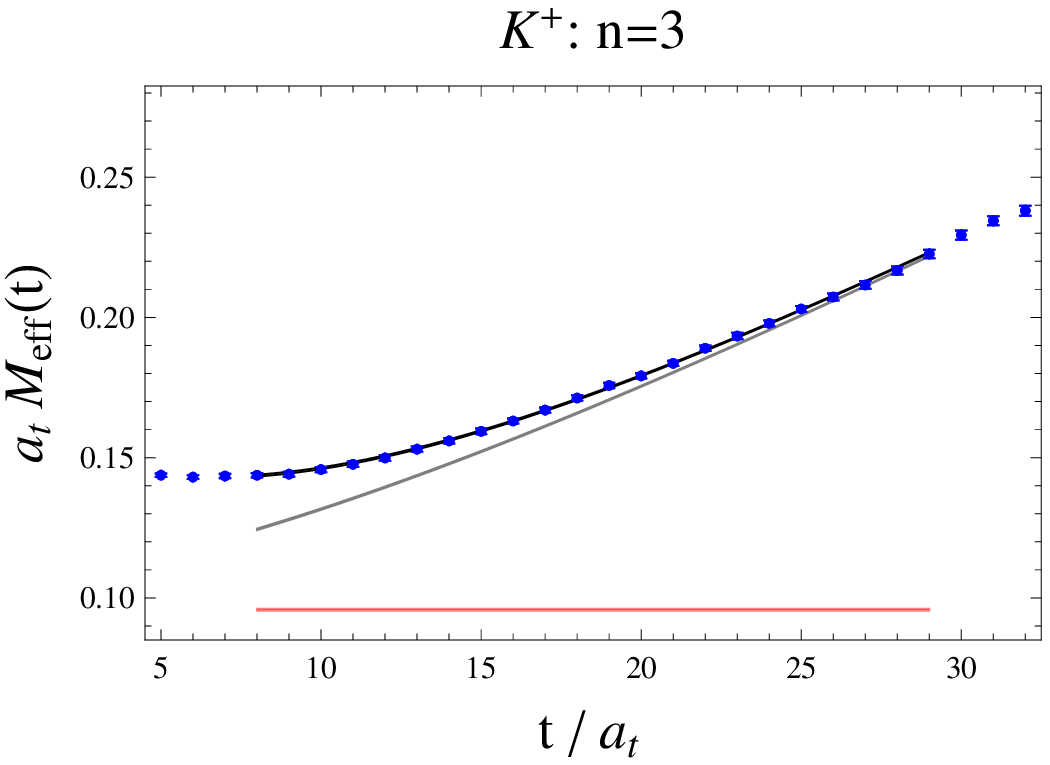,width=6.65cm}
\\
\epsfig{file=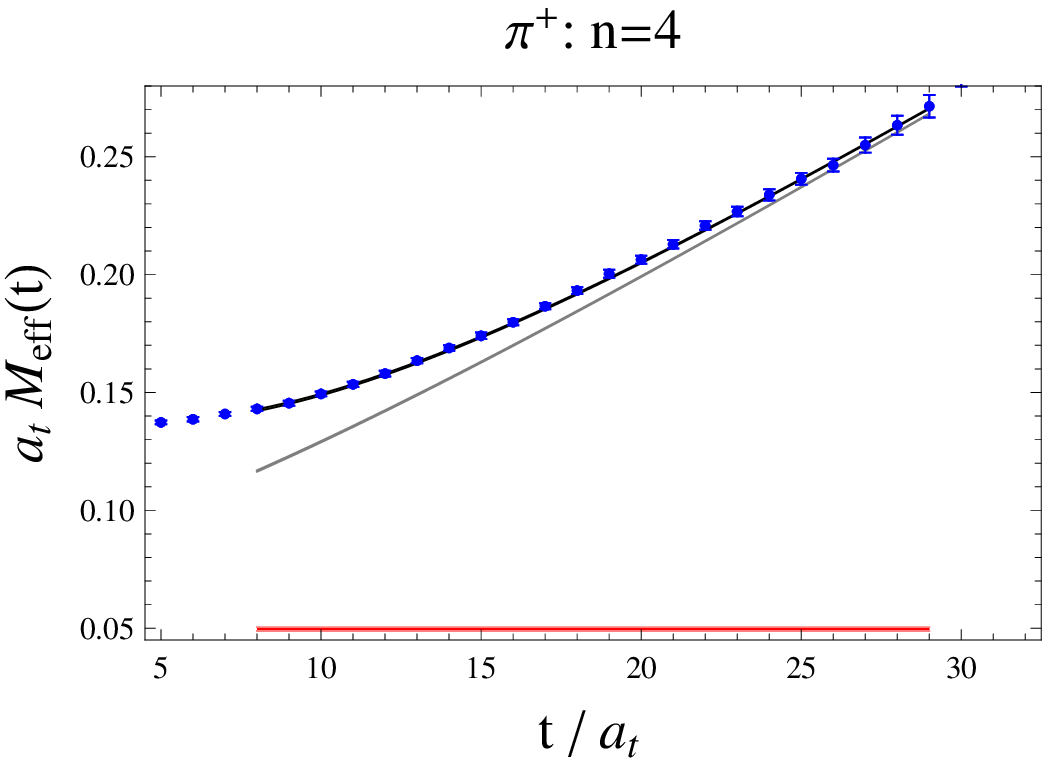,width=6.65cm}
$\phantom{sp}$
\epsfig{file=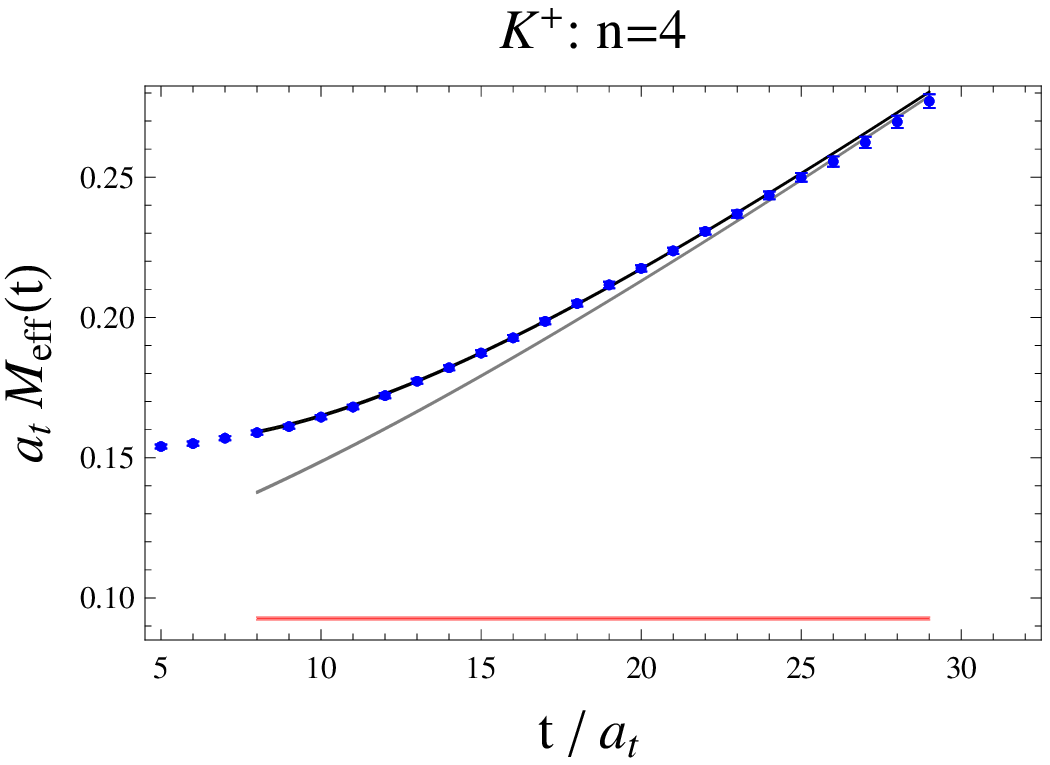,width=6.65cm}
\caption{
Charged pion and kaon effective mass plots  
along with two-state fits to the correlation function using
Eq.~\eqref{eq:GCharged}.
The bands represent the uncertainty from the extracted ground state 
rest energy, which is also plotted separately as a flat band. 
Using the fit parameters $Z(\cE)$ and $E(\cE)$, 
we form the ground state correlation function which is depicted by the intermediate (gray) bands.
Over long times, the effective mass should asymptotically approach these bands. 
We omit the zero field plots because they are quite similar to their uncharged 
counterparts. 
}
\label{f:PiPlusEeff}
\end{figure}

We utilize the 
conventional effective mass plot in order to display 
the non-standard behavior of charged particle 
correlation functions. 
In Fig.~\ref{f:PiPlusEeff}, 
we display effective mass plots for the charged pion and charged kaon. 
In non-vanishing fields, 
correlators exhibit a clear rise in the effective mass, 
Eq.~\eqref{eq:Meff}, 
with respect to time. 
The need for a fully relativistic treatment of the two-point function 
is also evident from the figure as effective-mass shifts are on the order of the 
rest mass.

Fits to the correlation functions of charged particles have been shown in the effective mass plots, 
Fig.~\ref{f:PiPlusEeff}. 
We fit the charged particle correlation functions 
using contributions from two states, 
as in Eq.~\eqref{eq:GCharged}
Although the amplitudes of the two states, 
$Z(\cE)$ 
and 
$Z'(\cE)$, 
enter the fit function linearly, 
we have not utilized  
variable projection
due to the increased computational time needed to perform the fits. 
In zero field, we augmented the correlation function 
with backwards propagating contributions to the two states. 
In non-vanishing electric fields, however, 
we found that backwards propagating charged particles make 
negligible contributions to the correlation functions.
For a $1\%$ effect due to a backwards propagating state, 
one must go beyond $t = 40$ for the $n=1$ field strength,
and to even larger times in stronger fields. 
Consequently we ignore backwards propagation in all 
but the zero-field case. 
Carrying out time-correlated fits on the bootstrap ensemble, 
we arrive at an ensemble of rest energies, 
$\{ \mathscr{E}_i(\cE) \}$, for the ground state.
At this point, the analysis parallels that of the neutral particles. 
Fits to the energy function are carried out on the bootstrap ensemble
using Eq.~\eqref{eq:Ecorr} producing the mass, polarizability, and quartic coupling. 
These extracted parameters are then averaged over the bootstrap ensemble.
Their uncertainties arise from both fitting and bootstrapping, 
which we have added in quadrature. 
%
%
%
%
%
\begin{table}[t]
\begin{center}
\begin{tabular}{cccc||cccc}
$$ & $n$ & $a_t E(\cE)$ & $1-P$ & $$ & $n$ & $a_t E(\cE)$ & $1-P$ 
\tabularnewline
\hline
\hline
$\quad \pi^+ \quad$ & $0$ & $0.0691(4) $ & $0.66$ &
$\quad K^+ \quad$  & $0$ & $0.0969(3) $ & $0.70 $ 
\tabularnewline
& $1$ & $0.0702(6) $ &$0.46$ &
& $1$ & $0.0979(4)$ & $0.77$
\tabularnewline
& $2$ & $0.0718(8) $ & $0.61$ &
& $2$ & $0.0982(7)$ & $0.78$
\tabularnewline
& $3$ & $0.0733(16) $ & $0.93$ &
& $3$ & $0.0958(10) $  & $0.98$
\tabularnewline
& $4$ & $0.0497(129) $ & $0.97$ &
& $4$ & $0.0927(23) $ & $0.97$ 
\tabularnewline
\hline
\hline
\tabularnewline
\end{tabular}
\begin{tabular}{ccccc||ccccc}
$\pi^+ $ & $\quad  a_t M \quad $ &  $\quad \alpha_E^{\text{latt}} \quad  $  & $\quad \ol \a^{\text{latt}}_{EEE} \quad $ & $1-P $ &
$K^+ $ & $\quad a_t M \quad $ & $\quad \alpha_E^{\text{latt}} \quad  $   & $\quad \ol \a^{\text{latt}}_{EEE} \quad $ & $ 1-P $
\tabularnewline
\hline
I & $0.0692(2)$ & $18(4)(6)$ & $24(10)$ & $0.30$ &
I & $0.0971(2)$ & $8(3)(1)$ & $17(5)$ &  $0.03$
\tabularnewline
II & $0.0692(2)$ & $16(3)(3)$ & $17(10)$ & $0.64$ &
II & $0.0969(2)$ & $16(4)(3)$ & $40(9)$ &  $0.23$
\tabularnewline
\hline
\hline
\end{tabular}
\end{center}
\caption{%
Summary of fit results for charged meson two-point functions for 
$8 \leq t \leq  29$.
Entries are as in Table~\ref{t:NeutralFit}.
}
\label{t:ChargedFit}
\end{table}
%
%
%
%

Extracted values of rest energies and fit parameters have been 
tabulated for the charged pion and kaon in Table~\ref{t:ChargedFit}. 
In performing these fits, 
we used the fit window $8 \leq t \leq 29$. 
By comparing fits on adjacent time windows, 
we can estimate the systematic due to the choice 
in fit window. 
We find a large spread in the extracted 
value of charged particle polarizabilities, 
and consequently a comparatively large systematic 
uncertainty due to the fit window. 
Rest energies are particularly sensitive to the fit window
as the field strength increases.

\begin{figure}[!t]
\epsfig{file=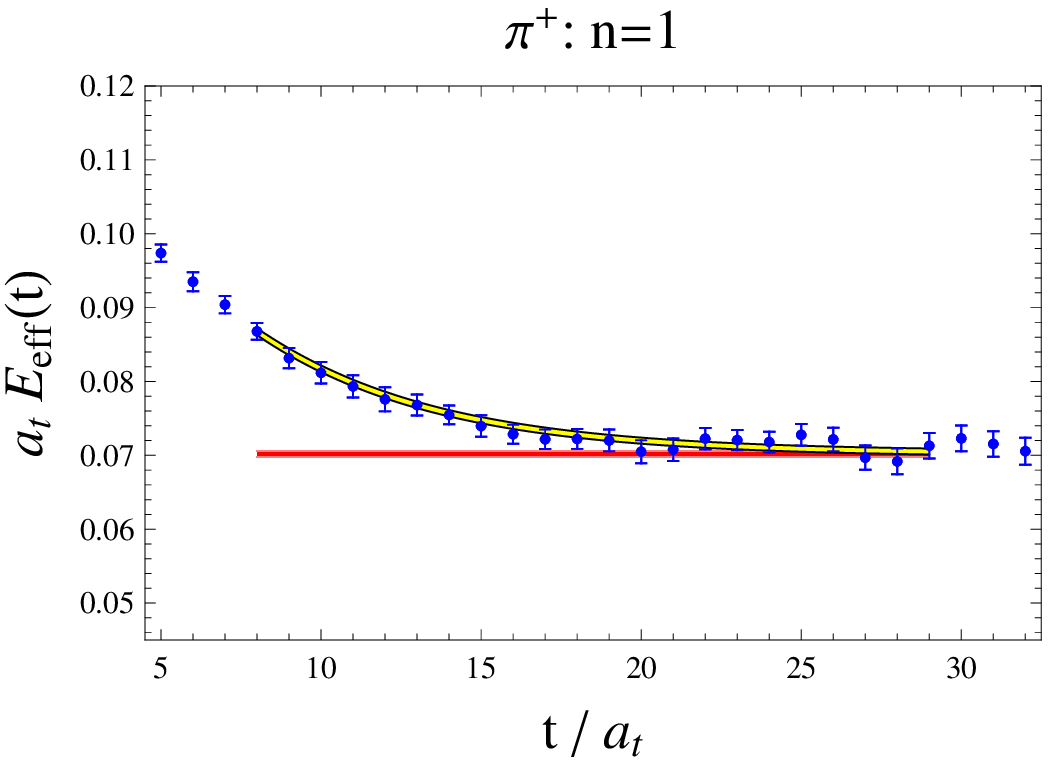,width=5.5cm}
$\phantom{sp}$
\epsfig{file=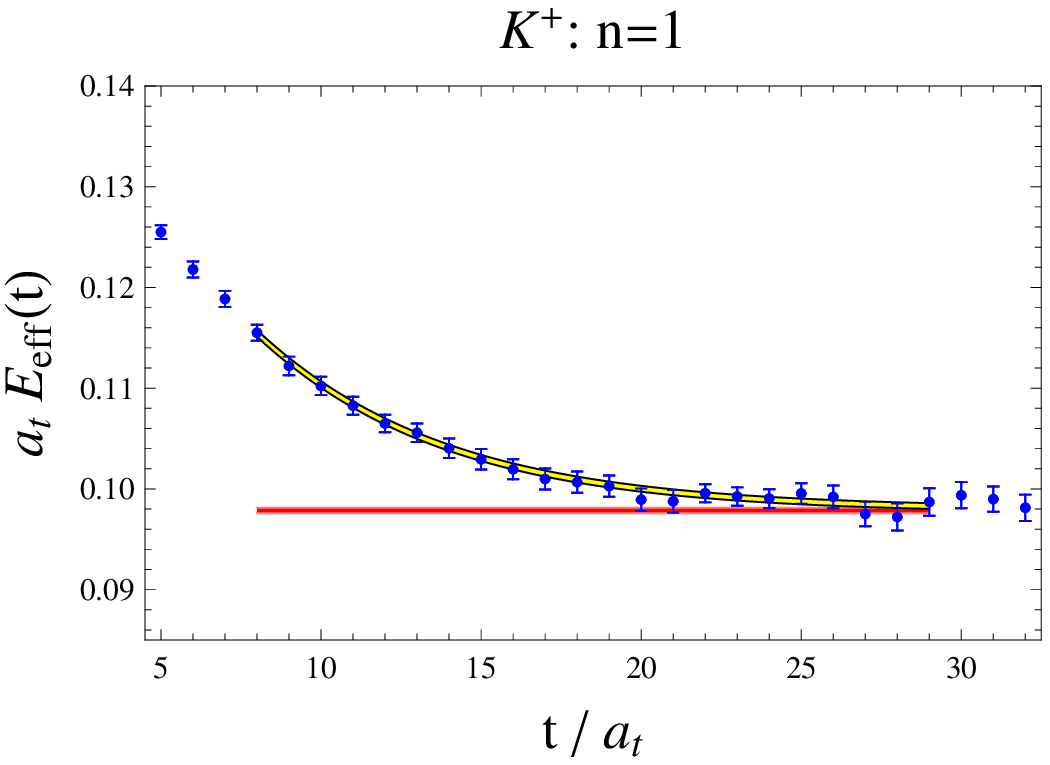,width=5.5cm}
\\
\epsfig{file=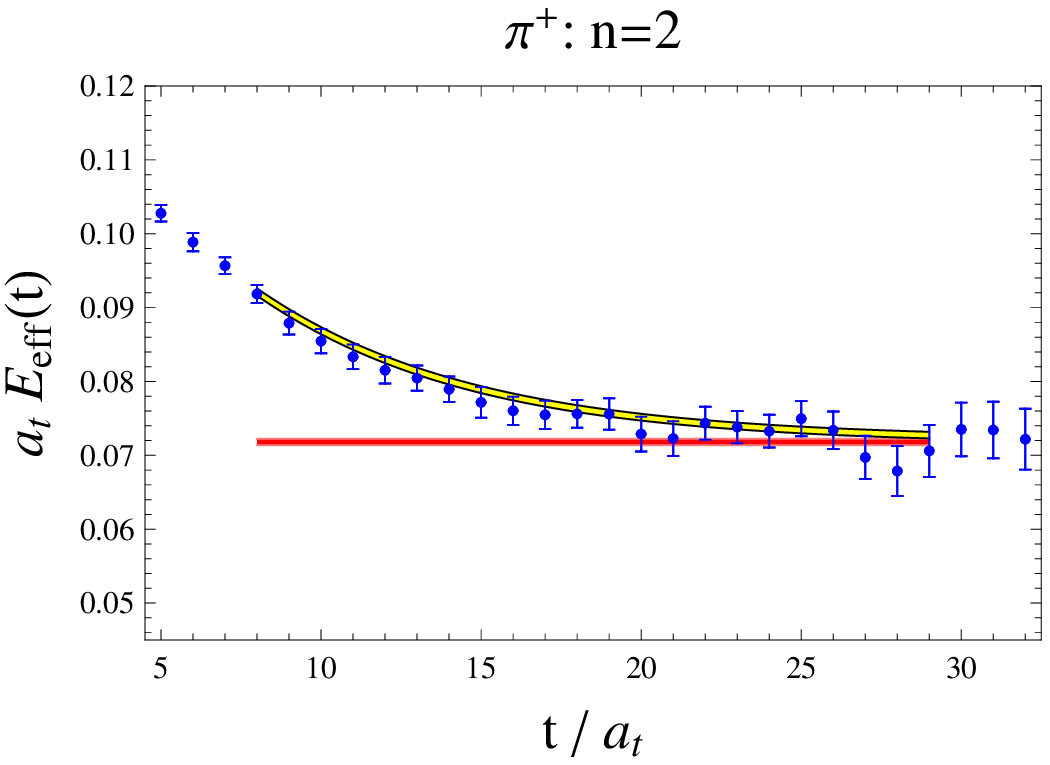,width=5.5cm}
$\phantom{sp}$
\epsfig{file=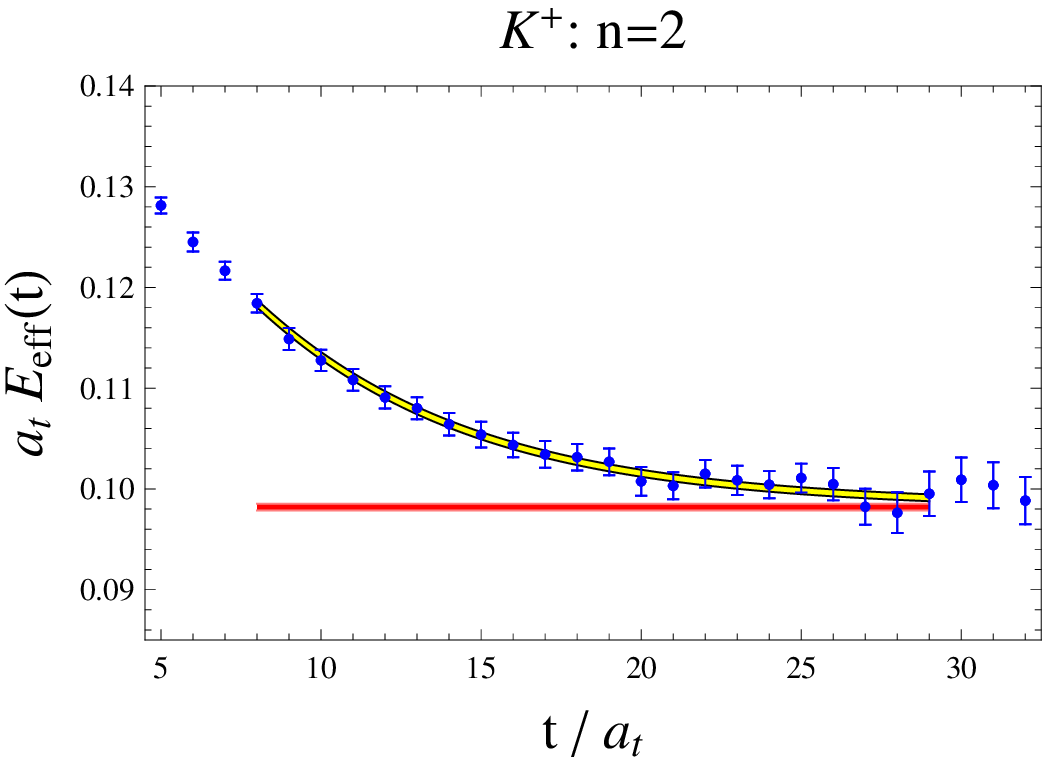,width=5.5cm}
\\
\epsfig{file=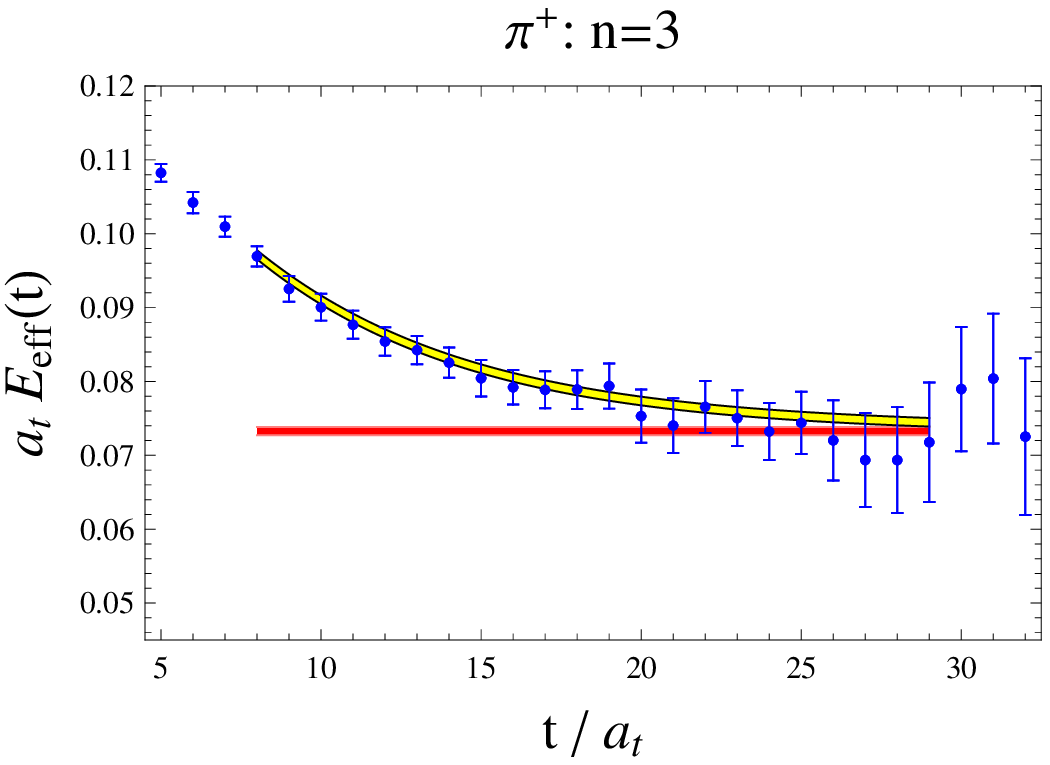,width=5.5cm}
$\phantom{sp}$
\epsfig{file=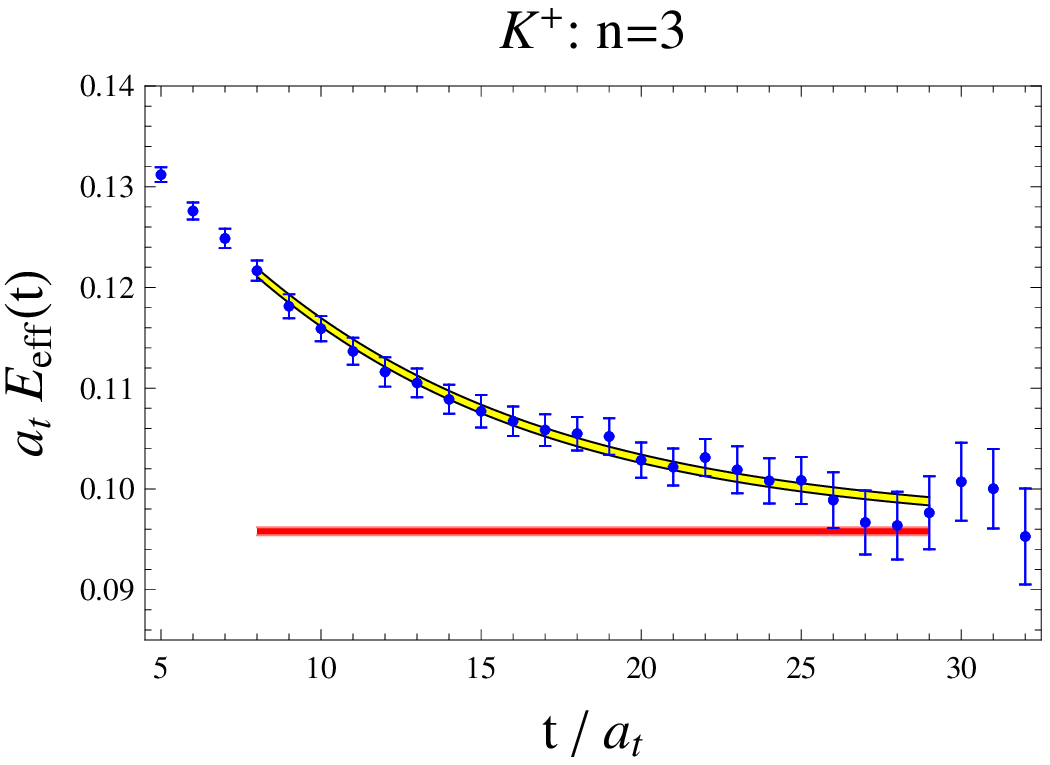,width=5.5cm}
\\
\epsfig{file=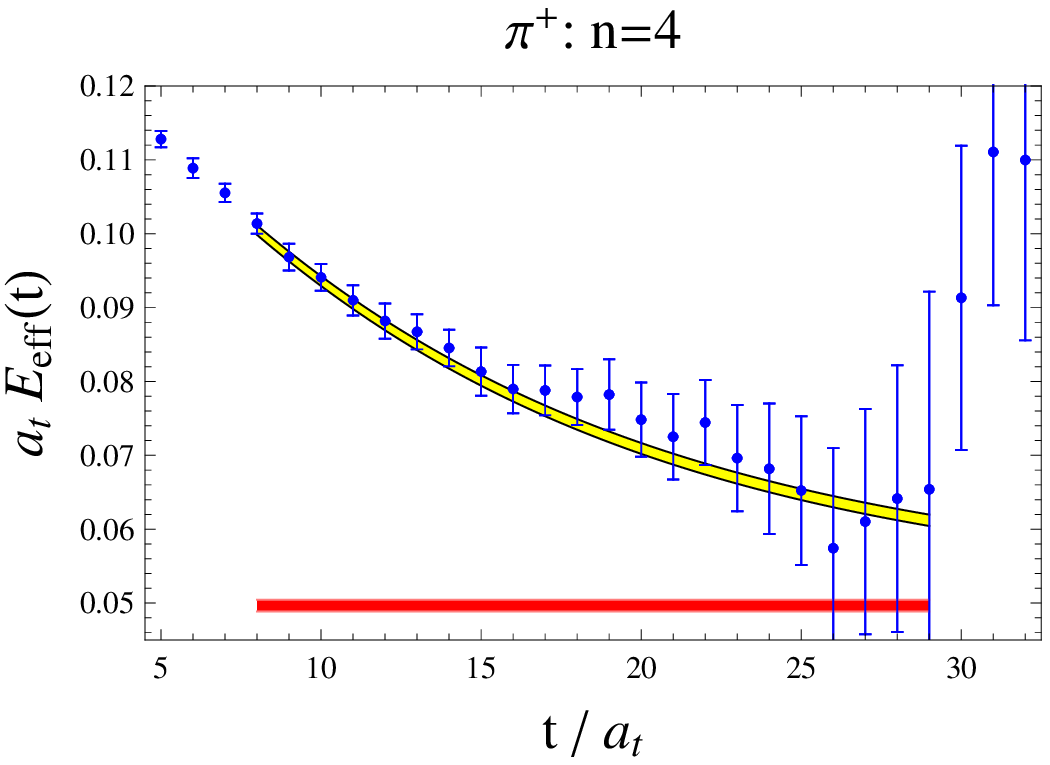,width=5.5cm}
$\phantom{sp}$
\epsfig{file=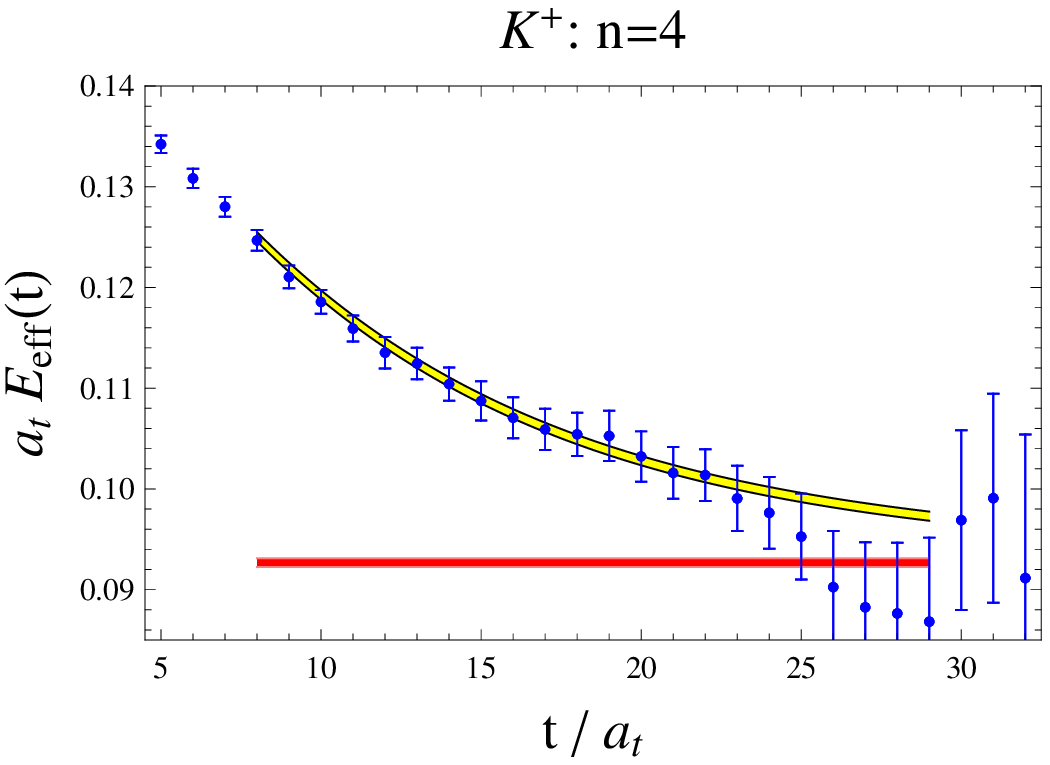,width=5.5cm}
\caption{
Effective energy plots for the charged pion and kaon two-point functions. 
Fits to the correlation functions are also shown.
Values for $n$ correspond to the strength of the quantized electric field $\cE$, 
given in Eq.~\eqref{eq:Quant}. 
We have omitted the $n=0$ plots, as they are just the ordinary effective mass plots.
}
\label{f:KPlusEeff}
\end{figure}

While fits to charged particle correlation functions appear to describe
the data well when displayed in terms of the effective mass,   
a further tool can be used to more clearly present these fits. 
This tool, moreover, aids in the determination of appropriate fit windows,
and we refer to it as the effective energy plot. 
The effective energy, just like the effective mass, is produced
by considering the correlation function at successive times. 
The relativistic propagator for a charged particle in Eq.~\eqref{eq:QProp}
depends on the time, the electric field, and rest energy,
$D = D\big(t, E(\cE), \cE\big)$,
albeit through a complicated one-dimensional integral. 
Given numerical data for the correlation function, 
$g(t,\cE)$, 
we can successively solve%
\footnote{
Because the effective energy is deduced from the non-linear
relation in Eq.~\eqref{eq:Eeff}, there is no guarantee a solution 
exists. 
Ensembles for which no solution can be found at a given time 
are dropped from the bootstrap.
This only affected error bars the 
$n=4$ 
effective energy plot for the $\pi^+$, 
and only for $t \geq 24$, 
where on average 
$5$ 
bootstraps were dropped.
}
for the effective energy in time by considering the ratio
\begin{equation} \label{eq:Eeff}
\frac{
D(t+1, E_{\text{eff}}, \cE) 
}
{
D(t,  E_{\text{eff}}, \cE )
}
=
\frac{g(t + 1,\cE)}{ g(t,\cE)}
,\end{equation}
with the value of the electric field, 
$\cE$, 
as input. 
This produces the effective energy as a function of time, 
$E_{\text{eff}} (t)$. 
Effective energy plots for the charged pion and kaon are shown 
in Fig.~\ref{f:KPlusEeff}. 
The effective energy should plateau over long times to the 
rest energy of the charged particle. 
From the figure, however, we see that contributions from the 
first excited state linger, and plateaus are not quite reached
before the noise grows substantially.
Nonetheless, 
we clearly see behavior reminiscent of the neutral particle effective mass plots in Fig.~\ref{f:PiZeroMeff}. 
This confirms that Eq.~\eqref{eq:QProp} properly describes the correlation function
of a charged particle in an electric field.

\begin{figure}[!t]
\epsfig{file=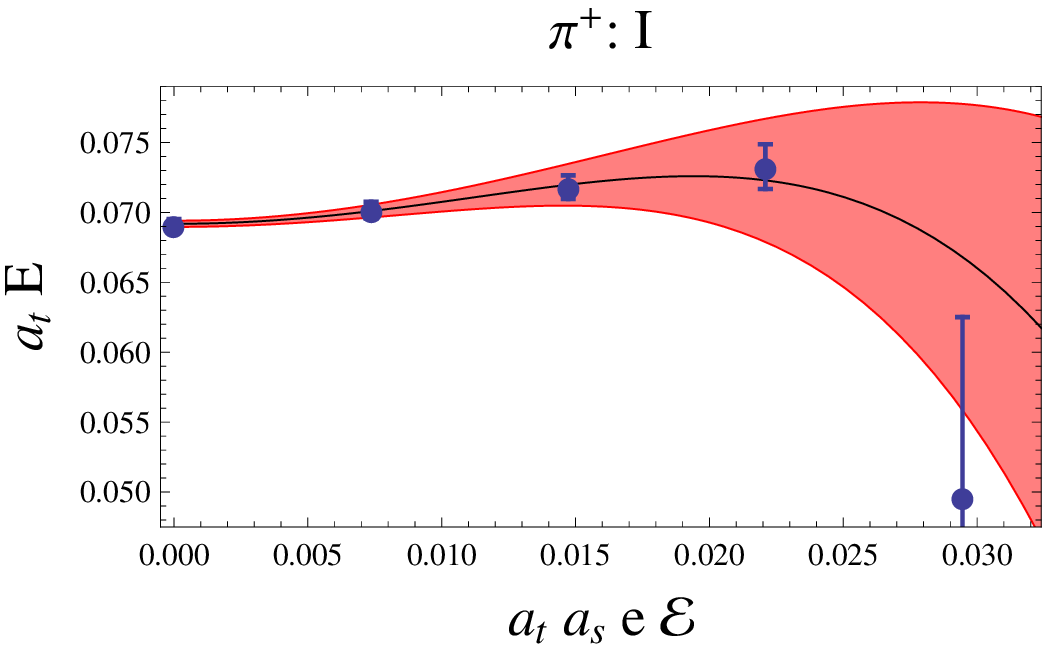,width=7.5cm}
\epsfig{file=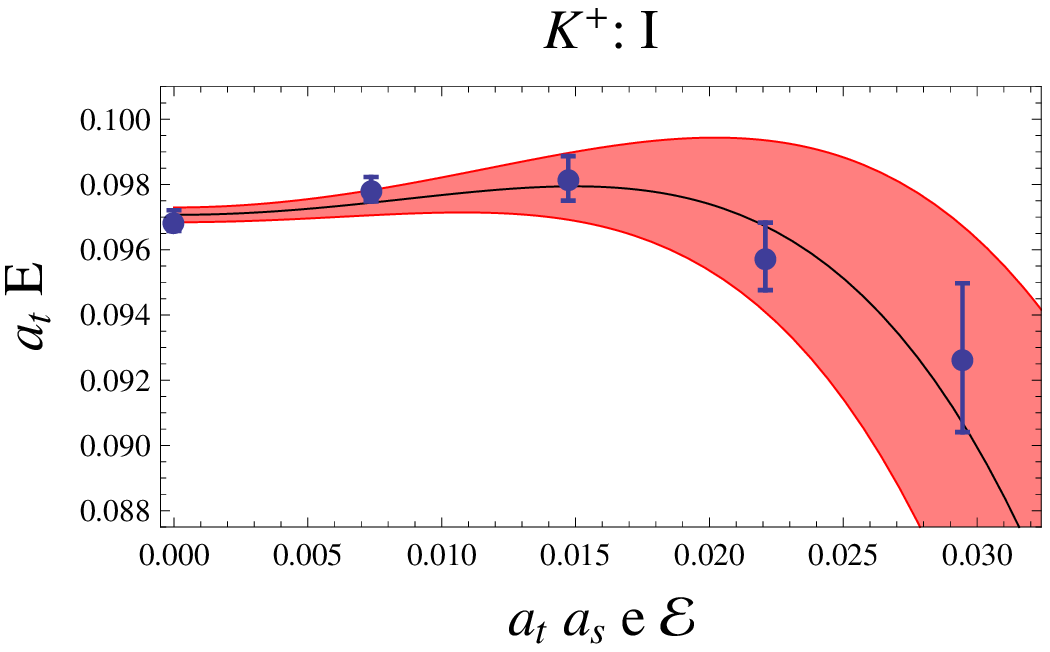,width=7.5cm}
\\
\epsfig{file=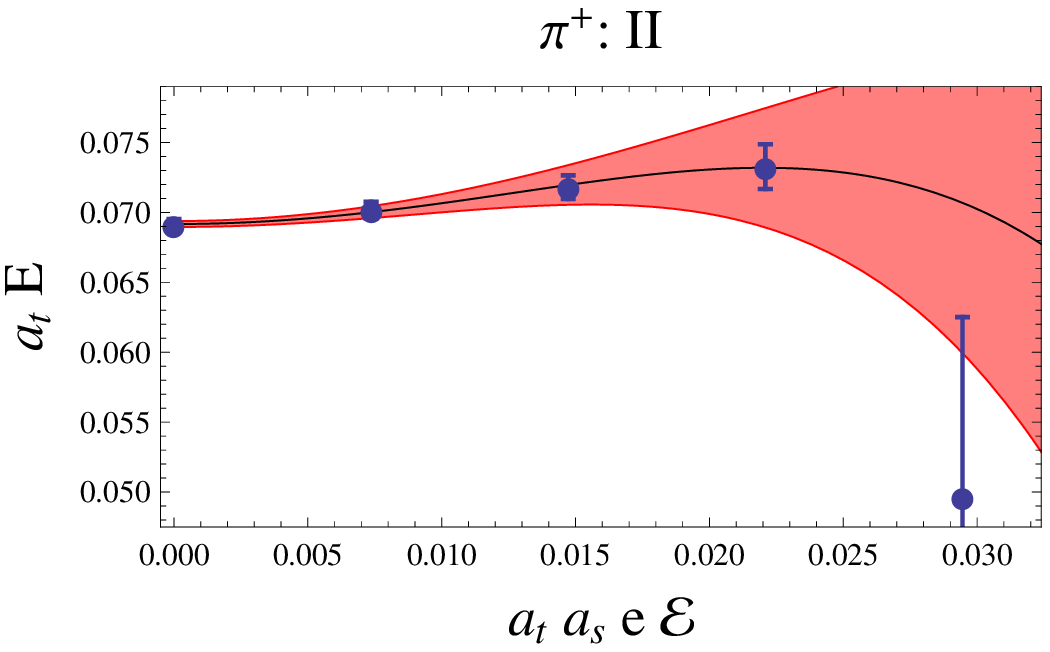,width=7.5cm}
\epsfig{file=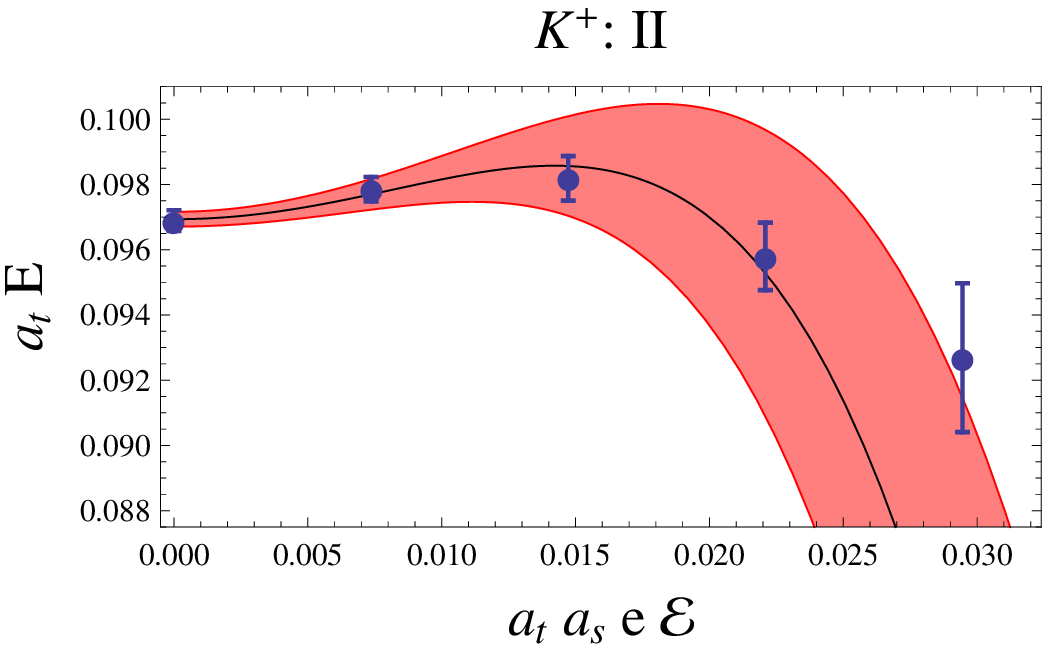,width=7.5cm}
\caption{
Plots of the electric field dependence of the extracted rest energies for charged mesons. 
}
\label{f:ChargedEvsE}
\end{figure}

Finally in Fig.~\ref{f:ChargedEvsE}, 
we plot the electric field dependence
of the extracted rest energies of the charged pion and kaon. 
There is striking non-monotonic behavior which indicates
the presence of quartic and perhaps higher-order terms in the field strength. 
We can make a brief comparison with chiral perturbation theory.
The size of the extracted polarizabilities is consistent with 
na\"ive expectations, 
i.e.~positive and on the order of $10^{-4} \texttt{fm}^3$ in physical units.

\section{Conclusion}                                                                                             %
\label{summy}                                                                                                         %

In this work, 
we have employed constant electric fields on a periodic lattice
to  investigate meson electric polarizabilities. 
Sizes of current-day lattices allow the utilization of 
properly quantized values of the electric field that lead to perturbative shifts
in hadron energies.
To test our setup, 
we have shown that the neutral pion (connected part) and kaon
polarizabilities can be extracted from lattice QCD
by measuring their energies as a function of the 
applied electric field strength.
Furthermore, 
we have investigated the charged pion and charged kaon polarizabilities, 
for which simple spectroscopy is of no avail. 
Using the relativistic charged particle propagator in the presence of an electric field, 
we fit lattice two-point functions and extract
rest energies of charged pions and kaons. 
Using effective energy plots, we showed that,
despite non-standard behavior for the correlation 
function, rest energies of charged particles show behavior
similar to the effective masses of neutral particles in electric fields. 
Charged meson polarizabilities were then extracted
from the behavior of the rest energy as a function of 
the electric field. 
Resulting electric polarizabilities have comparatively 
large uncertainties due predominantly to two sources.
With our current analysis method, 
the choice of time window gives a larger than expected
systematic uncertainty. 
Global fits, 
that are correlated in both time and electric field strength, 
can address such systematic error.
Secondly higher-order terms in the weak field expansion of 
charged particle rest energies appear to be very 
important,  
prompting future study on larger lattices on which 
the quantized field strengths are smaller. 
We hope that further refinements to the fitting procedure, 
additional data at different volumes and pion masses 
will remove the largest systematic effects, 
and ultimately bring lattice QCD in contact with experimental data for polarizabilities. 
Ultimately we will also use sea quarks that couple to the 
background fields.


\begin{acknowledgments}
These calculations were performed using 
the Chroma software suite~\cite{Edwards:2004sx}
on the computing clusters at Jefferson Laboratory.
Time on the clusters was awarded through the 
USQCD collaboration, and made possible by the SciDAC Initiative.
This work is supported in part by 
Jefferson Science Associates, LLC under 
U.S.~Dept.~of Energy contract No.~DE-AC05-06OR-23177 (W.D.).
The U.S. government retains a non-exclusive, paid-up
irrevocable, world-wide license to publish or reproduce
this mansuscript for U.S. government purposes. 
Additional support provided by the 
U.S.~Dept.~of Energy, under
Grant Nos.~DE-FG02-04ER-41302 (W.D.),
~DE-FG02-93ER-40762 (B.C.T.), and
~DE-FG02-07ER-41527 (A.W.-L.).
\end{acknowledgments}


\appendix

\section*{Appendix: Non-Uniform Fields}

Although we employ quantized field strengths, Eq.~\eqref{eq:Quant}, 
with a proper treatment of the boundary flux, Eq.~\eqref{eq:ModifiedAbelianLink}, 
we have additionally explored the effect of non-quantized fields on particle correlators. 
For this study, we use isotropic $24^3 \times 64$ lattices, 
the details of which are presented in~\cite{Detmold:2008xk}.
We summarize our findings here.

%
%
%
\begin{figure}
\includegraphics[width=7cm]{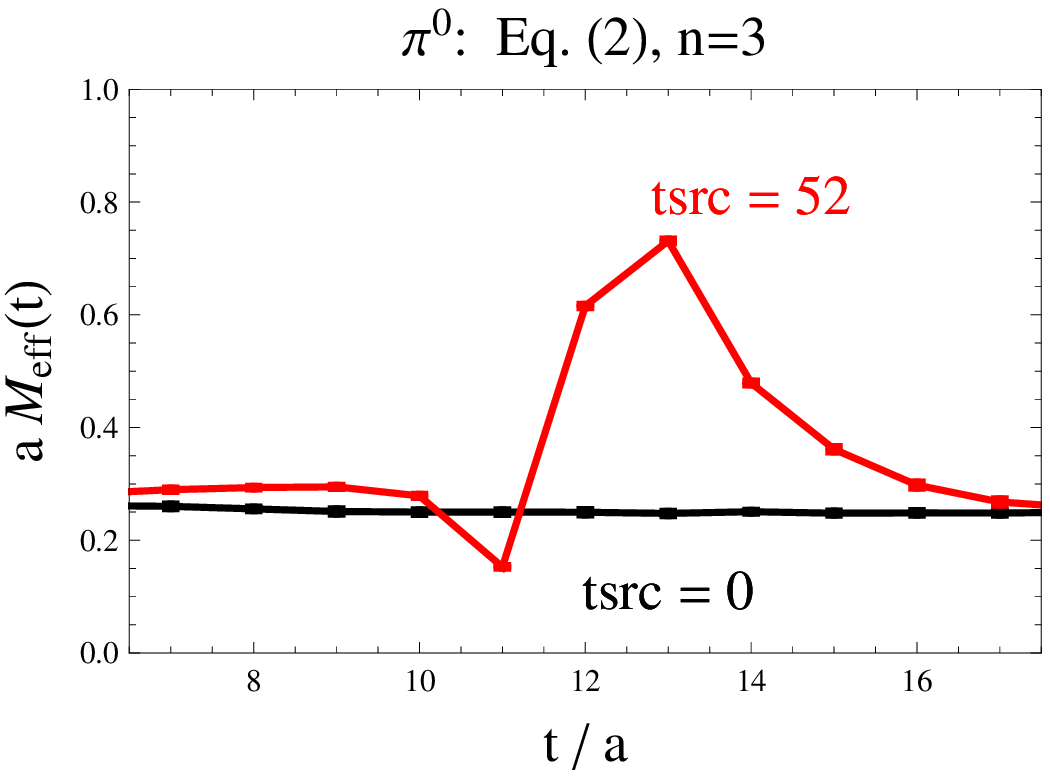}
$\qquad$
\includegraphics[width=7cm]{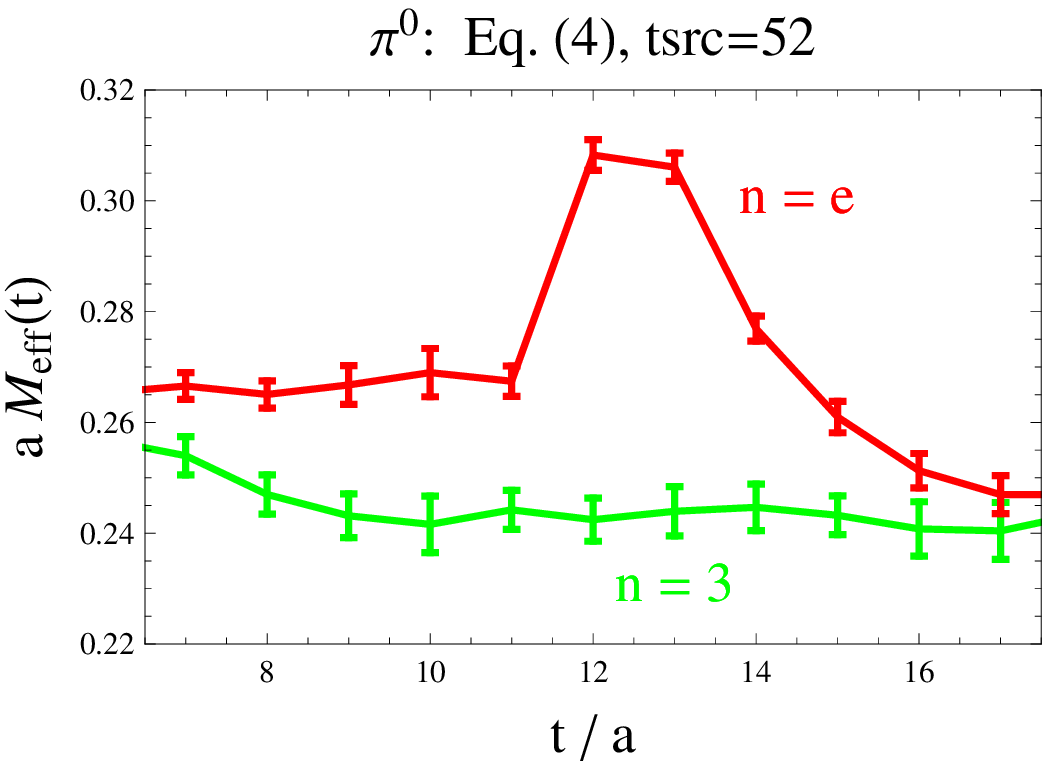}
\caption{\label{f:compare} 
Comparison of background field implementation.
On the left, we fix the field strength and change the source
location using Eq.~\eqref{eq:AbelianLink} to implement the background field. 
On the right, we fix the source time and change the field
strength while using Eq.~\eqref{eq:ModifiedAbelianLink}. 
}
\end{figure}
%
%
%
%

First we consider the na\"ive implementation of the external field using
Eq.~\eqref{eq:AbelianLink}
and the field value corresponding to 
$n=3$
in Eq.~\eqref{eq:Quant}.
In the continuum limit, the spike in the boundary flux contracts to a point
and the field becomes uniform. 
To test the uniformity of the field at finite lattice spacing, we look 
at the (connected) neutral pion two-point function. 
If we take the source time at 
$t_{\text{src}} = 0$, 
then the effective mass exhibits a plateau around 
$t = 12$, 
as shown in  Fig.~\ref{f:compare}. 
On the other hand, if we take the source time at
$t_{\text{src}} = 52$, 
then the plateau would set in as the pion wraps around the 
time boundary. 
The correlation function shows striking evidence for the
spike in the electric field from boundary flux.
Notice in plotting we have translated the latter correlation function forward by 
$12$ 
units in time.

Next we consider the proper implementation of the external field on 
a torus using Eq.~\eqref{eq:ModifiedAbelianLink}. 
We again consider the (connected) neutral pion two-point function. 
Fixing the source time at 
$t_{\text{src}} = 52$, 
we plot in  Fig.~\ref{f:compare} the resulting 
effective mass for two values of the field strength, 
$n =3$ 
and 
$n = e = 2.71828\ldots$ \,. 
We translate both correlation functions forward by 
$12$ 
units in time. 
For $n=e$, the effect of boundary flux has been mitigated (roughly by a factor of ten), 
but,
leads to easily measurable shifts in the particle energy. 
The quantized value, $n=3$, exhibits a plateau as the field is uniform across the 
time boundary.

\bibliography{bibfile}

\end{document}